\documentclass[showpacs,twocolumn,eqsecnum,prb,amsmath,amssymb]{revtex4}

\usepackage{graphicx}
\usepackage{dcolumn}
\usepackage{bm}
\begin{document}
\title{
Theory of Kondo lattices  and its application to \\
high-temperature
superconductivity and pseudo-gaps in cuprate oxides }
\author{Fusayoshi J. Ohkawa}
\affiliation{Division of Physics, Graduate School of
Science,  Hokkaido University, Sapporo 060-0810, Japan}
\email{fohkawa@phys.sci.hokudai.ac.jp}
\received{12 May 2003}
\begin{abstract}    
A theory of Kondo lattices is developed for the $t$-$J$ model on a square
lattice. The spin susceptibility  is described in a  form consistent with a
physical picture of Kondo lattices: Local spin fluctuations at different sites
interact with each other by a {\it bare} intersite exchange interaction, which
is mainly composed of two terms such as the superexchange interaction, which
arises from the virtual exchange of spin-channel pair excitations of electrons
across the Mott-Hubbard gap, and an exchange interaction arising from that of
Gutzwiller's quasi-particles. The {\it bare} exchange interaction is enhanced
by intersite spin fluctuations developed because of itself. The enhanced
exchange interaction is responsible for the development of superconducting
fluctuations as well as the Cooper pairing between Gutzwiller's
quasi-particles. On the basis of the microscopic theory, we develop a
phenomenological theory of {\it low}-temperature superconductivity and
pseudo-gaps in the under-doped region as well as high-temperature
superconductivity in the optimal-doped region.   Anisotropic pseudo-gaps open
mainly because of $d\gamma$-wave superconducting low-energy fluctuations:
Quasi-particle spectra around $(\pm\pi/a,0)$ and $(0,\pm\pi/a)$, with $a$ the
lattice constant, or  $X$ points at the chemical potential are swept away by
strong inelastic scatterings, and  quasi-particles are well defined only around
$(\pm\pi/2a,\pm\pi/2a)$ on the Fermi surface or line.  As temperatures decrease
in the vicinity of superconducting critical temperatures, pseudo-gaps become
smaller and the well-defined region is extending toward X points.  The
condensation of $d\gamma$-wave Cooper pairs eventually occurs at low enough
temperatures when the pair breaking by inelastic scatterings becomes small
enough. 
\end{abstract}
\maketitle

\section{Introduction}\label{SecIntroduction}

It is an important issue to elucidate the mechanism of
high-temperature (high-$T_c$) superconductivity occurring on CuO$_2$
planes.\cite{highTc,sc1,sc2,sc3,sc4} Cooper pairs can be bound by an
exchange interaction.\cite{Hirsch}  According to an early
theory,\cite{87SC1,87SC2}  one of the most possible mechanisms is the
condensation of $d\gamma$-wave Cooper pairs bound by the 
superexchange interaction.  Because the on-site Hubbard repulsion
is  strong in high-$T_c$ cuprate oxides, theoretical critical
temperatures $T_c$ for conventional Bardeen-Cooper-Schrieffer (BCS)
superconductivity or the condensation of isotropic $s$-wave Cooper
pairs must be very low. As long as the superexchange interaction is
antiferromagnetic and it works mainly between nearest neighbors,
theoretical $T_c$'s for the $d\gamma$ wave are much higher than
those for other waves. Many experiments imply that the condensation
of $d\gamma$-wave Cooper pairs occurs in the cuprate
oxides.\cite{Exp-d1,Exp-d2} 

The {\em hole}\cite{hole} doping for which observed $T_c$'s show a
maximum is called an optimal doping; dopings are classified into
over- and under-doped ones according to whether they are less or
more than the optimal one. Both of normal and superconducting states
in the optimal- and over-doped regions can be explained within the
theoretical framework of the early theory.\cite{87SC2} 
However, superconductivity in the under-doped region significantly
deviates from the prediction of the early theory. Observed $T_c$'s
decrease  with decreasing {\em hole} dopings, although theoretical
$T_c$'s never decrease.  Superconducting gaps $\epsilon_G(0)$ at
$T=0$~K keep increasing with decreasing {\em hole}
dopings.\cite{Shen2,Shen3,Ding,Ino,Renner,Ido1,Ido2,Ido4,Ekino}
Experimental ratios of $\epsilon_G(0)/k_BT_c$ are much large than
4.28 given by the  early theory.\cite{87SC2}  

Normal states above $T_c$ are anomalous in the under-doped
region; many experiments imply or show the opening of pseudo-gaps in
quasi-particle spectra. For example, the longitudinal nuclear
magnetic relaxation (NMR) rate $1/T_1T$ becomes smaller as
temperatures $T$ are lowered.\cite{yasuoka}
 According to angle resolved photoemission spectroscopy
(ARPES),\cite{Shen2,Shen3,Ding,Ino} quasi-particle spectra at the
chemical potential are swept away  around $(\pm\pi/a,0)$ and
$(0,\pm\pi/a)$, with $a$ the lattice constant. Scanning tunneling
spectroscopy (STS) directly shows the opening of
pseudo-gaps.\cite{Renner,Ido1,Ido2,Ido4,Ekino}  Pseudo-gaps start to
open around the so called mean-field critical temperature
defined by\cite{Ido2,Ido4}
$T_{c0}\simeq\epsilon_G(0)/4k_B$, which is much
higher than $T_c$ in the under-doped region. 

The anisotropy of pseudo-gaps deduced from ARPES is similar to that
of $d\gamma$-wave superconducting gaps. One may argue that the
opening of pseudo-gaps must be a precursor of that of
superconducting gaps. On the other hand, the development of
pseudo-gaps with decreasing $T$ is not monotonic in several STS
data;\cite{Ekino,Ido4} the magnitude of pseudo-gaps shows a maximum
at a temperature between $T_c$ and $T_{c0}$ and it decreases with
decreasing $T$ below the temperature. One may also argue that 
pseudo-gaps must be suppressed by superconductivity. It is one of
the most important issues to clarify how or whether the opening of
pseudo-gaps above $T_c$ is related with that of superconducting gaps
below $T_c$.

The cuprate oxides with no {\em hole} dopings are Mott-Hubbard's
insulators. Even if {\em holes} are doped, they must lie in the
strong-coupling regime defined by $U/W\agt 1$, with $U$ the Hubbard
repulsion and $W$ the bandwidth. The Hartree-Fock approximation, the
random-phase approximation (RPA) and their more or less improved
theories such as the self-consistent renormalization (SCR) theory
starting from RPA,\cite{Moriya} the fluctuation exchange (FLEX)
approximation  starting from RPA and so on are never any relevant
approximations for the cuprate oxides; they are relevant only for
weakly correlated electron liquids in the weak-coupling regime
defined by $U/W\ll 1$. 

Hubbard's\cite{Hubbard} and Gutzwiller's\cite{Gutzwiller} are within
the single-site approximation (SSA) and  are relevant for $U/W\agt
1$. When we take both of them,\cite{comHubGut} we can argue that 
the density of states must be of a three-peak structure,
Gutzwiller's quasi-particle band at the chemical potential between
the lower and upper Hubbard bands.  This speculation was confirmed
in another SSA theory;\cite{OhkawaSlave} a narrow  band, which is
nothing but Gutzwiller's band, appears at the top of the lower
Hubbard band when the electron density is less than half-filling. 
Gutzwiller's band is responsible for metallic behavior, and the
Mott-Hubbard splitting occurs in both metallic and insulating
phases. 

The validity of the SSA theories, Hubbard's and Gutzwiller's, 
implies that local fluctuations are responsible for the three-peak
structure.  Local fluctuations are rigorously considered in one of
the best SSA's that include all the single-site terms.\cite{SSAcom} Such an
SSA is reduced to determining and solving selfconsistently the Anderson
model,\cite{Mapping} which is one of the simplest effective
Hamiltonians for the Kondo problem. The Kondo problem has already
been
solved,\cite{singlet,poorman,Wilson,Nozieres,Yamada,Yosida,Exact} 
so that many useful results are available. One of the most essential
physics involved in the Kondo problem is that a localized magnetic
moment is quenched by local quantum spin fluctuations so that the
ground states is a singlet\cite{singlet} or a normal Fermi
liquid.\cite{Wilson,Nozieres}  The Kondo temperature
$T_{K}$ is defined as a temperature or  energy scale of local
quantum spin fluctuations.  The so called Abrikosov-Suhl or Kondo
peak in the Kondo problem corresponds to Gutzwiller's band. Their
bandwidth is of the order of $k_BT_K$, with $k_B$ the Boltzmann
constant. On the basis of the mapping to the Kondo problem, we 
argue that lattice systems must show a metal-insulator crossover as
a function of $T$: They are nondegenerate Fermi liquids at $T\gg
T_{K}$ because local thermal spin fluctuations are dominant, while
they are Landau's normal Fermi liquids at $T\ll T_{K}$ because
local quantum spin fluctuations are dominant and 
magnetic moments are quenched by them.
Local-moment magnetism occurs at $T\gg T_{K}$ while
itinerant-electron magnetism occurs at $T\ll T_{K}$. 
Superconductivity can only occur at $T\ll T_K$. 

The superexchange exchange interaction was originally derived for
Mott-Hubbard's
insulators.\cite{Kramers,AndersonSupEch,Goodenough,Kondo,Kanamori}
One may suspect that it must work only in insulating phases. On the
other hand, it was recently shown from a field theoretical
approach that it arises from the virtual exchange of spin-channel pair
excitation of electrons across the Mott-Hubbard
gap.\cite{OhkawaExc,OhkawaFerro} Gutzwiller's band plays no role in
this exchange process. It is unquestionable that the superexchange
interaction works even in metallic phases as long as the Mott-Hubbard
splitting exists. An exchange interaction arising from the virtual
exchange of spin-channel pair excitations of Gutzwiller's
quasi-particles also plays a role in metallic
phases.\cite{OhkawaExc,OhkawaFerro} 

A perturbative treatment of exchange interactions including the
above two ones starting from an {\it unperturbed} state constructed
in the SSA is nothing but a theory of Kondo lattices.  Because the
SSA is rigorous for Landau's normal Fermi liquids in infinite
dimensions,\cite{Metzner} it can also be formulated as a $1/d$
expansion theory, with $d$ the spatial dimensionality.  The theory
has already been applied to various phenomena occurring in strongly
correlated electron liquids, not only high-$T_c$
superconductivity\cite{OhkawaSC,OhkawaSC2} but also
itinerant-electron ferromagnetism,\cite{OhkawaFerro}, the
Curie-Weiss law of itinerant-electron magnets,\cite{Miyai}
field-induced ferromagnetism or metamagnetism,\cite{Satoh1} 
itinerant-electron antiferromagnetism,\cite{OhkawaAntiferro} and so
on.  The early theory\cite{87SC1,87SC2} of high-$T_c$
superconductivity is also within a first approximate framework of
the theory of  Kondo lattices. 

Because of anomalous properties of the under-doped region,  not a
few people speculate that a novel mechanism must be responsible for
high-$T_c$ superconductivity. On the other hand, we confirmed in
solving the Kondo problem that  adiabatic or analytical continuity
is one of the most important  concepts in
physics.\cite{Yamada,Yosida,idea} It is, of course, one of the basic
assumptions of Landau's Fermi-liquid theory. The theory of Kondo
lattices also relies on this concept. There is no phase transition
or no symmetry change between high-$T_c$ superconductivity in the
optimal-doped region and {\em low}-$T_c$ superconductivity in the
under-doped region. Then, analytical continuity tells us that the
theory of Kondo lattices, which can explain superconductivity in the
over- and optimal-doped regions, must apply to {\em low}-$T_c$
superconductivity in the under-doped region. One of the purposes of
this paper is to apply the theory of Kondo lattices to {\em
low}-$T_c$ superconductivity and pseudo-gaps in the under-doped
region.

This paper is organized as follows: In Sec.~\ref{prelim}, a theory
of Kondo lattices is developed for the $t$-$J$ model.  On the basis
of the microscopic theory developed in Sec.~\ref{prelim}, a
phenomenological theory of {\em low}-$T_c$ superconductivity and
pseudo-gaps is developed in Sec.~\ref{phenomenological}, where
several parameters are phenomenologically determined instead of
completing self-consistent procedures involved in the microscopic
theory. Discussions are given in Sec.~\ref{discussion}. Conclusions
are  given in Sec.~\ref{conclusion}.

\section{Theory of Kondo lattices}
\label{prelim}

We consider the $t$-$J$ model on a square lattice: 
%
\begin{eqnarray}
{\cal H} &=& 
- \sum_{ij\sigma} 
t_{ij} a_{i\sigma}^\dag a_{j\sigma} 
- \frac1{2} J \sum_{\left< ij\right>}
({\bf s}_i \cdot {\bf s}_j) 
\nonumber \\ &&  
+ U_{\infty} \sum_{i}
a_{i\uparrow}^\dagger a_{i\uparrow}
a_{i\downarrow}^\dagger a_{i\downarrow}  ,
\end{eqnarray}
with 
${\bf s}_i = \frac1{2}\sum_{\alpha\beta}  \left(
\sigma_x^{\alpha\beta}, \sigma_y^{\alpha\beta},
\sigma_z^{\alpha\beta} \right) a_{i\alpha}^\dagger a_{i\beta}$,
with $\sigma_x$, $\sigma_y$ and $\sigma_z$ the Pauli matrixes, and
the summation over $\left< ij\right>$  restricted to nearest
neighbors.  Here, notations are conventional.  The second term is
the superexchange interaction.  Because we are interested in
the cuprate oxides, we assume that
$J\!\simeq\! -(0.10$-0.15)\hspace{1pt}eV and the 
transfer integral between nearest-neighbors is $t_1\!\simeq\!
(0.4$-0.5)\hspace{1pt}eV: 
\begin{equation}\label{EqJ}
J/|t_1|\simeq -(0.2\mbox{--}0.3) . 
\end{equation}
The third term is introduced to exclude any doubly occupied sites,
so that $U_\infty/|t_1|\rightarrow +\infty$. Because no phase
transition is possible at nonzero temperatures in two dimensions, we 
assume that weak three dimensionality exists but the reduction of
$T_c$  due to it is small.

When Landau's normal Fermi liquid is considered, the
single-particle  self-energy  is  divided into
single-site and multi-site terms:
$\Sigma_\sigma(i\varepsilon_n,{\bf k})=
\tilde{\Sigma}_\sigma(i\varepsilon_n) +
\Delta\Sigma_\sigma(i\varepsilon_n,{\bf k})$.
The single-site term $\tilde{\Sigma}_\sigma(i\varepsilon_n)$ is
identical to the self-energy for a mapped Anderson model,  which
should be  self-consistently determined and solved.\cite{Mapping}  
The multi-site term is divided into two terms:
\begin{equation}\label{EqMultiSelf1}
\Delta\Sigma_\sigma(i\varepsilon_n,{\bf k}) =
\Delta\Sigma_\sigma({\bf k}) 
+ \Delta\Sigma_\sigma^\prime(i\varepsilon_n, {\bf k}) ,
\end{equation}
where the first term is independent of energies $i\varepsilon_n$
and 
$\Delta\Sigma_\sigma^\prime(i\varepsilon_n, {\bf k})\rightarrow0$ in
the limit of $|\varepsilon_n| \rightarrow +\infty$.

First, we construct an {\it unperturbed} state  in a renormalized
SSA, where not only $\tilde{\Sigma}_\sigma(i\varepsilon_n)$ but
also $\Delta\Sigma_\sigma({\bf k})$ are included. Then, the
single-particle Green function is described in such a way that
\begin{equation}
G_\sigma^{(0)}(i\varepsilon_n,{\bf k}) =
\frac1{i\varepsilon_n \!\!+\! \mu \!-\! E({\bf k}) 
\!-\! \tilde{\Sigma}_\sigma(i\varepsilon_n)
\!-\! \Delta\Sigma_\sigma({\bf k})} ,\!
\end{equation}
where $\mu$ is the chemical potential, and  
$E({\bf k}) = -2 \sum_{l} t_l \eta_{ls}({\bf k})$ is
the dispersion relation of electrons,
with $t_l$ the $l$th
nearest-neighbor one of the transfer integral $t_{ij}$ 
and $\eta_{ls}({\bf k})$ the form factor of the $l$th nearest
neighbors: 
\begin{subequations}
\begin{eqnarray}
\eta_{1s}({\bf k}) &=& \cos(k_xa) + \cos(k_ya), \\
\eta_{2s}({\bf k}) &=& \cos[(k_x+k_y)a] +
\cos[(k_x-k_y)a], \hspace{0.8cm}
\end{eqnarray}
\end{subequations}
and so on, with $a$ the lattice constant. 
The mapping condition to the Anderson model is given
by\cite{Mapping}
\begin{equation}\label{EqMapping}
\frac1{N} \sum_{\bf k} G_\sigma^{(0)}(i\varepsilon_n,{\bf k})
= \tilde{G}_\sigma(i\varepsilon_n) ,
\end{equation}
with $N$ the number of lattice sites and
$\tilde{G}_\sigma(i\varepsilon_n)$ the Green function of the 
Anderson model; the on-site interaction of the  Anderson model
should also be infinitely large.  

The single-site term is expanded in such a way that
\begin{equation}\label{EqExpansion}
\tilde{\Sigma}_\sigma(i\varepsilon_n)
= \tilde{\Sigma}_0 
\!\!+\! \bigl(1\!-\!\tilde{\phi}_\gamma\bigr)i\varepsilon_n
\!\!+\! \bigl(1\!-\!\tilde{\phi}_s\bigr)\! \frac1{2}g\mu_B H
\!+ \cdots \! ,
\end{equation}
in the presence of infinitesimally small fields $H$,
with $g$ being the $g$ factor and $\mu_B$ the Bohr magneton.
The expansion (\ref{EqExpansion}) is accurate for
$|\varepsilon_n|/k_BT_K \ll 1$; we approximately use it for
$|\varepsilon_n|/k_BT_K \alt 1$. The Wilson ratio, which appears in
the Kondo problem, is defined by
\begin{equation}
\tilde{W}_s = \tilde{\phi}_s/\tilde{\phi}_\gamma .
\end{equation}
Because charge fluctuations are suppressed, 
$\tilde{W}_s\rightarrow 2$ in the limit of half-filling or
$n\rightarrow 1$,\cite{Wilson,Yamada,Yosida} with $n$ the electron
density per site. Theories in SSA give approximate values of
$\tilde{\phi}_\gamma \simeq 1/|1-n]$.\cite{Gutzwiller,OhkawaSlave}
The dispersion relation of quasi-particles in the {\it unperturbed}
state is given by
\begin{equation}
\xi_0({\bf k}) = \frac1{\tilde{\phi}_\gamma}
\left[E({\bf k}) + \tilde{\Sigma}_0 
+ \Delta\Sigma_\sigma({\bf k}) - \mu \right], 
\end{equation}
and the Green functions are given by
$G_\sigma^{(0)}(i\varepsilon_n,{\bf k}) =
(1/\tilde{\phi}_\gamma)
g_\sigma^{(0)} (i\varepsilon_n, {\bf k})
+ \mbox{(incoherent part)} $,
with
\begin{equation}
g_\sigma^{(0)}(i\varepsilon_n,{\bf k})=
\frac1{i\varepsilon_n  -\xi_0({\bf k}) 
+ i \gamma\ \mbox{sgn}(\varepsilon_n) },
\end{equation}
being the coherent part describing quasi-particles;
the incoherent part describes the lower and upper Hubbard bands.
Here, a phenomenological life-time width $\gamma$ is
introduced; $\mbox{sgn}(x)$ is defined in such a way that
$\mbox{sgn}(x) =-1$ for $x<0$ while $\mbox{sgn}(x) =1$ for $x>0$.

When the irreducible polarization function in spin channels 
is denoted by $\pi_s(i\omega_l, {\bf q})$, the susceptibility of
the $t$-$J$ model, which does not include the conventional factor
$\frac1{4}g^2\mu_B^2$, is given by
\begin{equation}\label{EqSus-0}
\chi_s(i\omega_l, {\bf q}) =
\frac{2\pi_s(i\omega_l, {\bf q})}
{
1- \left[ \frac1{2}J({\bf q})+ U_\infty \right]
\pi_s(i\omega_l, {\bf q})}, 
\end{equation}
with
\begin{equation}\label{EqSuperJ}
J({\bf q}) = 2J \eta_{1s}({\bf q}).
\end{equation}
The irreducible polarization function is also divided into
single-site and multi-site terms:
$\pi_s(i\omega_l, {\bf q}) = \tilde{\pi}_s(i\omega_l) + 
\Delta\pi_s(i\omega_l, {\bf q}) $.
The single-site term $\tilde{\pi}_s(i\omega_l)$ is identical
to that of the Anderson model, so that the susceptibility
of the Anderson model is given by
\begin{equation}
\tilde{\chi}_s(i\omega_l) =
\frac{2\tilde{\pi}_s(i\omega_l)}{1-U_\infty\tilde{\pi}_s(i\omega_l)} .
\end{equation}
 An energy scale of local quantum spin fluctuations  or the Kondo
temperature is defined by
\begin{equation}\label{EqDefTK}
k_BT_K =1/\bigl[\tilde{\chi}_s(0) \bigr]_{T \rightarrow +0} .
\end{equation}
Because $U_\infty|\tilde{\chi}_s(i\omega_l)|\rightarrow +\infty$ and
$U_\infty|\chi_s(i\omega_l, {\bf q})|\rightarrow +\infty$, 
Eq.~(\ref{EqSus-0}) can also be written in another form: 
\begin{equation}\label{EqKondoSus}
\chi_s(i\omega_l, {\bf q}) =
\frac{\tilde{\chi}_s(i\omega_l)} 
{1 - \frac1{4}I_s(i\omega_l, {\bf q})
\tilde{\chi}_s(i\omega_l)} ,
\end{equation}
with
\begin{equation}\label{EqIs}
I_s (i\omega_l, {\bf q}) = J({\bf q}) + 
2 U_\infty^2 \Delta\pi_s(i\omega_l, {\bf q}) .
\end{equation}
This expression is consistent with a physical picture of Kondo
lattices: Local spin fluctuations at different sites interact with
each other by an intersite exchange interaction. Then, we call
$I_s(i\omega_l, {\bf q})$ an exchange interaction.

According to the Ward relation,\cite{ward} the irreducible
single-site three-point vertex function in spin channels is given by
$\tilde{\lambda}_s(i\varepsilon_n,
i\varepsilon_n \!+\! i\omega_l;i\omega_l)
 =
\tilde{\phi}_s
\left[ 1 \!-\! U_\infty \tilde{\pi}_s (i\omega_l) \right]$ or
\begin{equation}\label{EqThreeL}
\tilde{\lambda}_s(i\varepsilon_n,
i\varepsilon_n+i\omega_l;i\omega_l)
= 2\tilde{\phi}_s / U_\infty\tilde{\chi}_s(i\omega_l)  ,
\end{equation}
for $\varepsilon_n/k_BT_K \rightarrow +0$ and 
$\omega_l/k_BT_K \rightarrow +0$. We approximately use
Eq.~(\ref{EqThreeL}) for nonzero 
$\varepsilon_n$ and $\omega_l$. Then, the so called
spin-fluctuation mediated interaction in the longitudinal spin
channel is given by 
\begin{equation}\label{EqIs*1}
\left[
U_\infty \tilde{\lambda}_s (0,0;0)
\right]^2 \!
\left[ \chi_s(i\omega_l, {\bf q}) \!-\! 
\tilde{\chi}_s(i\omega_l) \right] = 
\tilde{\phi}_s^2 
\frac1{4} I_s^* (i\omega_l, {\bf q}) ,
\end{equation}
with
\begin{subequations}\label{EqIs*2}
\begin{eqnarray}\label{EqIs*2A}
\frac1{4} I_s^*(i\omega_l, {\bf q}) &=&
\frac{ \frac1{4}
I_s (i\omega_l, {\bf q}) }
{1 - \frac1{4}I_s(i\omega_l, {\bf q})
\tilde{\chi}_s(i\omega_l) } 
\\ &=& 
\frac1{4} I_s(i\omega_l, {\bf q}) +
\left[ 
\frac1{4} I_s(i\omega_l, {\bf q})\right]^2
\!\! \chi_s(i\omega_l, {\bf q}).
\nonumber \\  && \label{EqIs*2B}
\end{eqnarray}
\end{subequations}
In Eq.~(\ref{EqIs*1}), the single-site term is subtracted because it is
considered in SSA.  The interaction in the transversal channels is
also given by Eq.~(\ref{EqIs*1}); the spin space is isotropic in
our system. Because of Eqs.~(\ref{EqIs*1}) and (\ref{EqIs*2}), we
call $I_s(i\omega_l, {\bf q})$ a {\em bare} exchange
interaction, $I_s^*(i\omega_l, {\bf q})$ an enhanced one, and
$\tilde{\phi}_s$ an effective three-point vertex function.

The {\em bare} exchange interaction (\ref{EqIs}) is
mainly composed of three terms:
\begin{equation}\label{EqIs1}
I_s(i\omega_l,{\bf q}) = J({\bf q}) + 
J_Q(i\omega_l,{\bf q}) - 4\Lambda (i\omega_l,{\bf q}) .
\end{equation}
The first term is the superexchange interaction (\ref{EqSuperJ}).
The second term is an exchange interaction arising from the virtual
exchange of pair excitations of quasi-particles. 
When Eq.~(\ref{EqThreeL}) is made use of, its lowest-order
term in intersite processes is given by\cite{OhkawaExc}
\begin{equation}\label{EqJQ}
\frac1{4} J_Q(i\omega_l,{\bf q}) = 
\frac{\tilde{W}_s^2}{\tilde{\chi}_s^2(0)} \bigl[
P(i\omega_l,{\bf q})-P_0(i\omega_l)\bigr] ,
\end{equation}
with
\begin{eqnarray}\label{EqP}
P(i\omega_l,{\bf q}) &=& - k_B T \sum_{\varepsilon_n}
\frac1{N} \sum_{{\bf k}\sigma}
g_\sigma^{(0)}(i\varepsilon_n+i\omega_l, {\bf k}+{\bf q})
\nonumber \\ && \qquad \times 
g_\sigma^{(0)}(i\varepsilon_n, {\bf k}) .
\end{eqnarray}
The summation over $\varepsilon_n$ can be analytically carried out
as is shown in Eq.~(\ref{EqPA}).
Here, the single-site term,
$P_0(i\omega_l) =  (1/N) \sum_{\bf q} P(i\omega_l,{\bf q})$, is
subtracted because it is considered in SSA. 
The third term corresponds to the so called mode-mode
coupling term in the SCR theory.\cite{Moriya} 
Because the $\omega$-linear imaginary term in $P_0(\omega+i0)$ is
cancelled by that in $1/\tilde{\chi}_s(\omega+i0)$,\cite{OhkawaExc} 
Eq.~(\ref{EqKondoSus}) is approximately given by 
\begin{equation}\label{EqKondoSus2}
\chi_s(\omega+i0, {\bf q}) =
\frac1 
{1/\tilde{\chi}_s(0) - \frac1{4}I_s^\prime(\omega+i0, {\bf q})} ,
\end{equation}
for $|\omega|\alt k_BT_K$; the exchange interaction is given by
\begin{subequations}
\begin{eqnarray}\label{EqJQ2}
&&\hspace*{-1.5cm} 
I_s^\prime(\omega \!+\! i0,{\bf q}) = J({\bf q}) + 
J_Q^\prime(\omega \!+\! i0,{\bf q}) - 4 \Lambda(0,{\bf Q}),
\\ &&\hspace*{-1.5cm} 
\frac1{4} J_Q^\prime(\omega \!+\! i0,{\bf q}) = 
\frac{\tilde{W}_s^2}{\tilde{\chi}_s^2(0)} \bigl[
P(\omega \!+\! i0,{\bf q})-P_0(0)\bigr] ,
\end{eqnarray}
\end{subequations}
with ${\bf Q} = (\pm\pi/a, \pm\pi/a)$.

As is shown in  Eq.~(\ref{EqMultiSelf1}), there are two types of
multi-site self-energy corrections. One is the Fock term due to the
superexchange interaction. When only the coherent part is
considered, it is calculated in such a way that
\begin{eqnarray}\label{EqSelfAFh}
\frac1{ \tilde{\phi}_\gamma} 
\Delta \hspace{-1pt}\Sigma_\sigma ({\bf k})  &=&
\frac{3}{4}\tilde{W}_s^2 \frac{k_B T}{N} \!
\sum_{\varepsilon_{n}{\bf q}}
 J({\bf q})  e^{i \varepsilon_{n}0^+}\!
g_\sigma^{(0)} (i \varepsilon_{n}, {\bf k}\!+\!{\bf q}) 
\nonumber \\ &=& 
2 J \Xi \eta_{1s}({\bf k}),
\end{eqnarray}
with
\begin{equation}
\Xi = \frac{3}{4} \tilde{W}_s^2 \frac{k_B T}{N} \!
\sum_{\varepsilon_{n}{\bf p}}
\eta_{1s}({\bf p}) e^{i \varepsilon_{n}0^+}\!
g_\sigma^{(0)} (i \varepsilon_{n}, {\bf p}) .
\end{equation}
Here,  the factor 3 appears because of three spin channels. 
According to Gutzwiller's,\cite{Gutzwiller}
$\tilde{\phi}_\gamma\rightarrow +\infty$ in the limit of
$n\rightarrow 1$ and the bandwidth of quasi-particles vanishes. 
When the Fock term is included, however, the bandwidth is about
$|J|$ in the {\em unperturbed} state even in the limit of
$n\rightarrow 1$, if there is no disorder. 

\begin{figure}
\centerline{
\includegraphics[width=9.0cm]{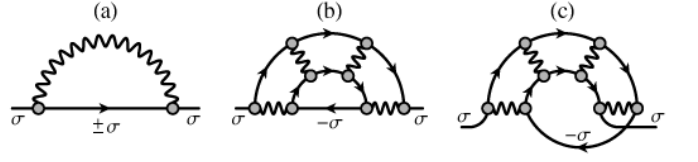}
}
\caption[1]{ 
Feynman diagrams for the multi-site self-energy. A solid lines
stands for  $g_\sigma^{(0)}(i\varepsilon_n,{\bf k})$, a wavy line for
$I_s^*(i\omega_l,{\bf q})$, and a solid circle for
$\tilde{W}_s$. Diagram (a) gives the sum of
$\Delta\Sigma_\sigma ({\bf k})$ and
$\Delta \Sigma_\sigma^{(AF)} (i\varepsilon_n,{\bf k})$. All the
orders of two types of
diagrams such as (b) and (c) are considered for 
$\Delta \Sigma_\sigma^{(SC)} (i\varepsilon_n,{\bf k})$. 
}
\label{feyman}
\end{figure}

So far the {\em unperturbed} state is constructed in a
renormalized SSA where $\tilde{\Sigma}_\sigma(i\varepsilon_n)$ and 
$\Delta\Sigma_\sigma ({\bf k})$ should be self-consistently
calculated.  Next, 
$\Delta \Sigma_\sigma^\prime (i\varepsilon_n,{\bf k})$ in
Eq.~(\ref{EqMultiSelf1}) is perturbatively considered in terms of
$I_s (i\omega_l, {\bf q})$ or $I_s^*(i\omega_l, {\bf q})$. We
consider two types of corrections shown in Fig.~\ref{feyman}:
$\Delta \Sigma_\sigma^\prime (i\varepsilon_n,{\bf k}) =\Delta
\Sigma_\sigma^{(AF)} (i\varepsilon_n,{\bf k}) +\Delta
\Sigma_\sigma^{(SC)} (i\varepsilon_n,{\bf k})$. One arises from
antiferromagnetic spin fluctuations:
\begin{eqnarray}\label{EqSelfAF}
\frac1{\tilde{\phi}_\gamma}
\Delta \Sigma_\sigma^{(AF)} (i\varepsilon_n,{\bf k}) &=&
\frac{k_B T}{N}  \sum_{\omega_{l}{\bf q}}
\tilde{W}_s^2 U_{AF}(i\omega_l,{\bf q}) 
\nonumber \\ && \times 
g_\sigma^{(0)} 
(i \varepsilon_{n}\!+\!i\omega_l, {\bf k}\!+\!{\bf q}),
\qquad 
\end{eqnarray}
with
\begin{eqnarray}\label{EqUAF}
U_{AF}(i\omega_l,{\bf q}) &=&  
\frac{3}{4} \Bigl[ I_s^*(i\omega_l,{\bf q})-J({\bf q})
\Bigr] .
\end{eqnarray}
Here, the superexchange interaction $J({\bf q})$, which gives the
Fock term, is subtracted.
%

The other arises from $d\gamma$-wave superconducting fluctuations.
We define a reducible polarization function of $d\gamma$-wave
particle-particle or Cooper-pair channel by 
\begin{eqnarray}
\Pi_{\sigma\sigma^\prime}(i\omega_l, {\bf q}) &=&
\frac1{N} \sum_{\bf pk} 
\eta_{d\gamma}\left({\bf p}\right)
\eta_{d\gamma}\left({\bf k}\right)\!
\int_{0}^{1/k_BT} \hspace{-15pt} d \tau e^{i \omega_l \tau}
\nonumber \\ && \hspace*{-2.5cm} \times
\left<
a_{({\bf p}+\frac1{2}{\bf q}) \sigma}(\tau)
a_{(-{\bf p}+\frac1{2}{\bf q}) -\sigma}(\tau)
a_{(-{\bf k}+\frac1{2}{\bf q}) -\sigma^\prime}^{\dagger}
a_{({\bf k}+\frac1{2}{\bf q}) \sigma^\prime}^{\dagger}
\right> ,
\nonumber \\ &&
\end{eqnarray}
with
\begin{equation}
\eta_{d\gamma}({\bf k}) = \cos(k_xa) - \cos(k_ya).
\end{equation}
Cooper
pairs are bound with the enhanced exchange interaction
 (\ref{EqIs*2}).
It is expanded in such a way that
\begin{equation}\label{EqPairInt}
I_s^*(i\omega_l, {\bf q}) =
I_0^*(i\omega_l)
+2 \sum_{i} I_{i}^*(i\omega_l) \eta_{is} ({\bf q}) .
\end{equation}
Because the nearest-neighbor $I_1^*(i\omega_l)$
plays a major role, we only consider it and we ignore
other ones. When only ladder diagrams such as those
shown in Figs.~\ref{feyman}(b) and (c) are considered, fluctuations
of different waves are decoupled from each other, so that
$\Pi_{\sigma\sigma^\prime}(i\omega_l, {\bf q}) $ satisfies the
following coupled equation:
\begin{subequations}\label{EqBS}
\begin{eqnarray}
&& \hspace*{-0.5cm}
\Pi_{\sigma\sigma}(i\omega_l, {\bf q}) \!=\!
\pi_{d\gamma}(i\omega_l, {\bf q}) \!+\!
\frac1{2} \pi_{d\gamma}(i\omega_l, {\bf q}) I_1^*\tilde{\phi}_s^2
\Pi_{\sigma-\sigma}(i\omega_l, {\bf q})
\nonumber \\ && \hspace*{1.5cm}
- \frac1{4}\pi_{d\gamma}(i\omega_l, {\bf q}) I_1^*\tilde{\phi}_s^2
\Pi_{\sigma\sigma}(i\omega_l, {\bf q}) ,
\\ &&
\Pi_{\sigma-\sigma}(i\omega_l, {\bf q}) =
\frac1{2} \pi_{d\gamma}(i\omega_l, {\bf q}) I_1^* \tilde{\phi}_s^2
\Pi_{\sigma\sigma}(i\omega_l, {\bf q}) 
\nonumber \\ && \hspace*{1.5cm}
- \frac1{4} \pi_{d\gamma}(i\omega_l, {\bf q}) I_1^* \tilde{\phi}_s^2
\Pi_{\sigma-\sigma}(i\omega_l, {\bf q}) . 
\end{eqnarray}
\end{subequations}
Here,  relations of 
$\Pi_{\sigma-\sigma}(i\omega_l, {\bf q}) 
=\Pi_{-\sigma\sigma}(i\omega_l, {\bf q})$ and
$\Pi_{\sigma\sigma}(i\omega_l, {\bf q}) 
=\Pi_{-\sigma-\sigma}(i\omega_l, {\bf q})$ as well as 
\begin{eqnarray}
2 \eta_{1}({\bf k}-{\bf p}) &=& 
\eta_{1s}({\bf k})\eta_{1s}({\bf p})
+ \eta_{d\gamma}({\bf k})\eta_{d\gamma}({\bf p})
\nonumber \\ && 
+ \eta_{x}({\bf k})\eta_{x}({\bf p})
+ \eta_{y}({\bf k})\eta_{y}({\bf p}), \hspace{1cm}
\end{eqnarray}
with $\eta_{x}({\bf k})=\sqrt{2}\sin (k_xa)$ and
$\eta_{y}({\bf k})=\sqrt{2}\sin (k_ya)$, are made use of; the energy
dependence of $I_1^*(i\omega_l)$ is ignored and 
$I_1^*=\left<\mbox{Re}[I_1^*(\omega+i0)]\right>$ is an average over
a low-energy region such as $|\omega|\alt k_BT_K$; 
 $\pi_{d\gamma}(i\omega_l, {\bf q})=
\pi_{d\gamma}^*(i\omega_l, {\bf q})/\tilde{\phi}_\gamma^2$, with
\begin{eqnarray}\label{EqIrredPiSC}
\pi_{d\gamma}^*(i\omega_l, {\bf q}) &=&
- \frac{k_BT}{N} \sum_{\varepsilon_n{\bf k}}
\eta_{d\gamma}^2\left({\bf k}\right) 
g_\sigma^{(0)}(i\varepsilon_n,{\bf k}+
\mbox{$\frac1{2}{\bf q}$})
\nonumber \\ && \quad\times 
g_\sigma^{(0)}(-i\varepsilon_n-i\omega_l,
-{\bf k}+\mbox{$\frac1{2}{\bf q}$}) .
\qquad 
\end{eqnarray}
The summation over $\varepsilon_n$ can be analytically carried out
as is shown in Eq.~(\ref{EqIrredPiSCA}).
As is discussed in Sec.~\ref{SecIntroduction}, $T_c$'s for the
$d\gamma$ wave are much higher than $T_c$'s for other waves, so that
low-energy fluctuations of the $d\gamma$ wave dominate over those of
other waves.  Then, we consider only them. A superconducting
susceptibility for the
$d\gamma$ wave is given  by
\begin{eqnarray}\label{EqChiSC}
\chi_{d\gamma}^*(i\omega_l, {\bf q}) &\equiv&
\tilde{\phi}_s^2 \left[
\Pi_{\sigma\sigma}(i\omega_l, {\bf q}) 
-\Pi_{\sigma-\sigma}(i\omega_l, {\bf q}) \right]
\nonumber \\ &=&
\frac{\displaystyle 
\tilde{W}_s^2 \pi_{d\gamma}^*(i\omega_l, {\bf q})}
{
1 + \frac{3}{4}I_1^*  
\tilde{W}_s^2 \pi_{d\gamma}^*(i\omega_l, {\bf q})  } .
\end{eqnarray}
The self-energy correction  is given by
\begin{eqnarray}\label{EqSelfSC}
\frac1{\tilde{\phi}_\gamma }
\Delta \Sigma_\sigma^{(SC)} (i\varepsilon_n,{\bf k}) &=&
- \frac{k_B T}{N} 
\sum_{\omega_{l}{\bf q}}  
\tilde{W}_s^2 \eta_{d\gamma}^2 ({\bf k}+\mbox{$\frac1{2}$}{\bf q})
\nonumber \\ && \hspace*{-2cm} \times 
U_{d\gamma}(i\omega_l, {\bf q}) 
g_\sigma^{(0)}(-i\varepsilon_{n}\!-\!i\omega_l, -{\bf k}\!-\!{\bf q}),
\qquad \quad 
\end{eqnarray}
with 
\begin{equation}\label{EqUdG}
U_{d\gamma}(i\omega_l, {\bf q}) =
\left(\frac{3}{4} I_1^*  \right)^2  
\chi_{d\gamma}^*(i\omega_l, {\bf q}) .
\end{equation}
In Eqs.~(\ref{EqChiSC}) and (\ref{EqUdG}), the factor 3 appears
together with $I_1^*$ because of three spin channels.

The  Green function renormalized  by
antiferromagnetic and superconducting fluctuations is given by 
$G_\sigma(i\varepsilon_{n}, {\bf k}) =
g_\sigma (i\varepsilon_{n}, {\bf k})/\tilde{\phi}_\gamma
+ \mbox{(incoherent part)}$,
with
\begin{equation}\label{EqG*2}
g_\sigma (i\varepsilon_{n}, {\bf k}) =
\left[ 
\frac1{g_\sigma^{(0)} (i \varepsilon_{n}, {\bf k})}
- \frac1{\tilde{\phi}_\gamma}
\Delta\Sigma_\sigma^\prime(i \varepsilon_{n}, {\bf k})
\right]^{-1} \!\!, 
\end{equation}
with $\Delta\Sigma_\sigma^\prime(i \varepsilon_{n}, {\bf k})
=\Delta\Sigma_\sigma^{(AF)}(i \varepsilon_{n}, {\bf k})
+\Delta\Sigma_\sigma^{(SC)}(i \varepsilon_{n}, {\bf k})$.
The density of states for quasi-particles  is
given by
\begin{equation}
\rho^*(\varepsilon) = \frac1{N} \sum_{\bf k}
\rho_{\bf k}^*(\varepsilon) ,
\end{equation}
with
\begin{equation}
\rho_{\bf k}^*(\varepsilon) =
\mbox{Im} \left( - \frac1{\pi} \right)
g_\sigma(\varepsilon+i0,{\bf k}) .
\end{equation}

\section{Semi-phenomenological theory}
\label{phenomenological}
\subsection{High-$T_c$ superconductivity}
\label{high-Tc}

On the basis of the formulation in Sec.~\ref{prelim}, we develop a
phenomenological theory.  First, we consider the {\em unperturbed}
state.  The dispersion relation $\xi_0({\bf k})$ of quasi-particles
in the {\em unperturbed} state is expanded as
\begin{equation}
\xi_0({\bf k}) = 
\mu^* - \sum_{i} 2 t_i^*\eta_{is}({\bf k})  . 
\end{equation}
Although the expansion coefficients such as $\mu^*$, $t_1^*$,
$t_2^*$, and so on should be self-consistently determined, we treat
them as phenomenological parameters. We assume that
\begin{equation}\label{t1*}
t_1^*>0 , \quad  
t_2^* = - 0.3 t_1^*, 
\end{equation}
and we ignore other $t_l^*$'s. Eq.~(\ref{t1*}) is
consistent with experiment.  Furthermore, we treat $t_1^*$ as an
energy unit for the sake of simplicity, although it
depends on temperatures, electron densities and so on. 

The density of states for quasi-particles in the {\it
unperturbed} state is given by
\begin{equation}
\rho_\gamma (\varepsilon) = \frac1{N} \sum_{\bf k}
\left(-\frac1{\pi} \right) \mbox{Im} g_\sigma^{(0)}
(\varepsilon+i0,{\bf k}) . 
\end{equation}
According to the Fermi-surface sum rule,\cite{FSrule} the electron
density is given by that of quasi-particles, so that in
the limit of $T\rightarrow +0$ and $\gamma\rightarrow +0$
\begin{eqnarray}\label{EqFSumRule}
n &=& 
2\frac{k_BT}{N}\sum_{\varepsilon_n{\bf k}}
 e^{i \varepsilon_{n}0^+} 
g_\sigma^{(0)} (i \varepsilon_{n}, {\bf k})
\nonumber \\ &=& 
2\!\! \int \!\! d\varepsilon \rho_{\gamma=0}(\varepsilon) 
f_\gamma(\varepsilon) 
%
= 2\!\! \int \!\! d\varepsilon \rho_{\gamma}(\varepsilon) 
f_{\gamma=0}(\varepsilon) , \quad
\end{eqnarray}
with
\begin{equation}
f_\gamma(\varepsilon)  =
\frac1{2} +\frac1{\pi} \mbox{Im}  \left[
\psi\left( \frac1{2} 
+ \frac{\gamma - i\varepsilon}{2\pi k_B T} \right)
\right] .
\end{equation}
with $\psi (z)$ the di-gamma function.
Note that
$\rho_{\gamma=0}(\varepsilon) = (1/N)\sum_{\bf k}
\delta[\varepsilon - \xi_0({\bf k})]$ and
$f_{\gamma=0}(\varepsilon) =1/[e^{\varepsilon/k_BT} + 1] $.

\begin{figure*}
\centerline{
\includegraphics[width=18.5cm]{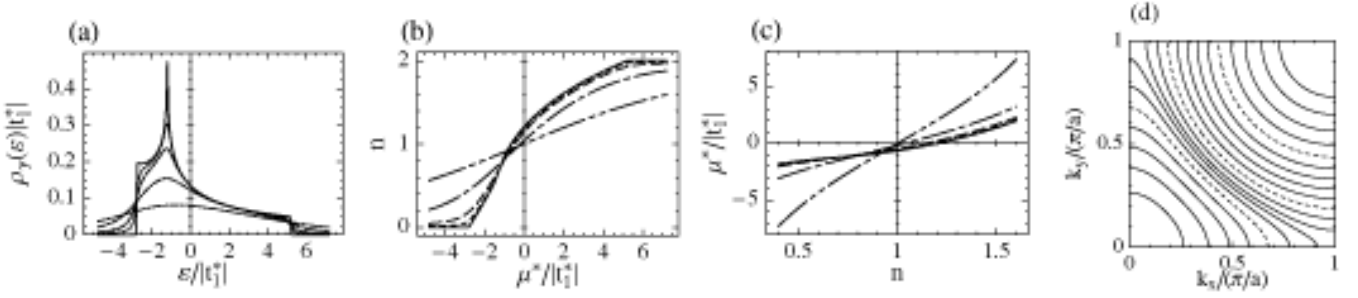}}
\caption[2]{
Single-particle properties of the {\it unperturbed} state for
$T=0$~K, $t_1^* >0$ and $t_2^* = - 0.3t_1^*$.  
(a) Density of states for quasi-particles
$\rho_\gamma(\varepsilon)$,  
(b) the electron density $n$ as a function of the effective
chemical potential $\mu^*$, and  
(c) $\mu^*$ as functions of $n$.
In these three figures, solid, dotted, broken, dot-broken and
two-dot-broken lines are for
$\gamma/|t_1^*|=10^{-3}$, 0.1, 0.3, 1 and 3, respectively. 
(d) Fermi surfaces for $\gamma=0$ and 19 electron densities such as  
$n=0.1 \times i$, with $1\le i \le 19$ being an integer. Dotted lines
show those for $n=$0.5, 1.0 and 1.5.
More than half-filling cases $(n > 1)$ can be treated, if
we take the hole\cite{hole} picture. }
\label{rho_n_mu_FS}
\end{figure*}

We assume Eq.~(\ref{EqFSumRule}) even for nonzero $T$ and $\gamma$.
Nonzero $\gamma$'s are introduced, partly, for the convenience of
numerical processes.  Fig.~\ref{rho_n_mu_FS}(a)
shows $\rho_\gamma (\varepsilon)$ for several $\gamma$. As long as
$\gamma$'s are rather small such as
$\gamma/|t_1^*|\alt 0.3$, we expect that essential features are
never wiped out by such $\gamma$.  
 Fig.~\ref{rho_n_mu_FS}(b) shows
$n$ as a function of $\mu^*$,  Fig.~\ref{rho_n_mu_FS}(c) shows
$\mu^*$ as a function of $n$, and Fig.~\ref{rho_n_mu_FS}(d) shows
Fermi surfaces for $\gamma=0$ and several $n$.

\begin{figure*}
\centerline{\hspace*{0.5cm}
\includegraphics[width=18.5cm]{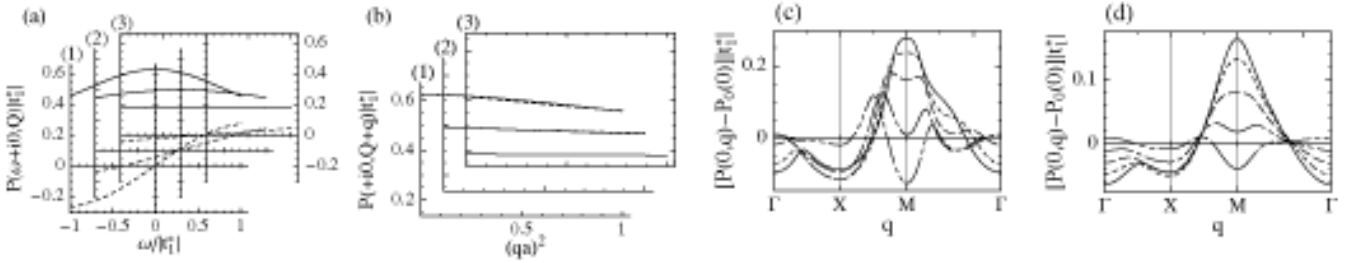}
} 
\caption[3]{ 
Spin-channel polarization functions: 
(a) $P(\omega+i0,{\bf Q})|t_1^*|$ and
(b) $P(+i0,{\bf Q}+{\bf q})|t_1^*|$, with 
${\bf Q}=(\pm\pi/2,\pm\pi/2)$, for  $n=0.9$ and
$k_BT/|t_1^*|=0.1$; 
$\gamma/|t_1^*|=0.3$, 1 and
3 are assumed, respectively, in  (1), (2) and (3).
In Fig.~\ref{Fig_P}(a), solid and dotted
lines show the real and imaginary parts, respectively; 
In Fig.~\ref{Fig_P}(b), solid and dotted lines show those for 
${\bf q}\parallel (1,0)$ and ${\bf q} \parallel (1,1)$,
respectively;  the polarization functions are almost isotropic
around {\bf Q}. (c) $[P(+i0,{\bf q})-P_0(+i0)]|t_1^*|$
for $k_BT/|t_1^*|=\gamma/|t_1^*|=0.1$, and
(d) $[P(+i0,{\bf q})-P_0(+i0)]|t_1^*|$
for  $k_BT/|t_1^*|=\gamma/|t_1^*|=0.3$; solid, dotted, broken,
dot-broken and two-dot-broken lines are for
$n=1$, 0.9, 0.8, 0.7 and 0.6, respectively.
Because $P(+i0,{\bf q})$'s have peaks
at wave-numbers different from ${\bf Q}$ for small $n$, effective
$\kappa_s$'s had better been estimated from Figs.~\ref{Fig_P}(c) and
(d) instead of  Fig.~\ref{Fig_P}(b).
 }
\label{Fig_P}
\end{figure*}  
\begin{figure*}
\centerline{\hspace*{0.2cm}
\includegraphics[width=18cm]{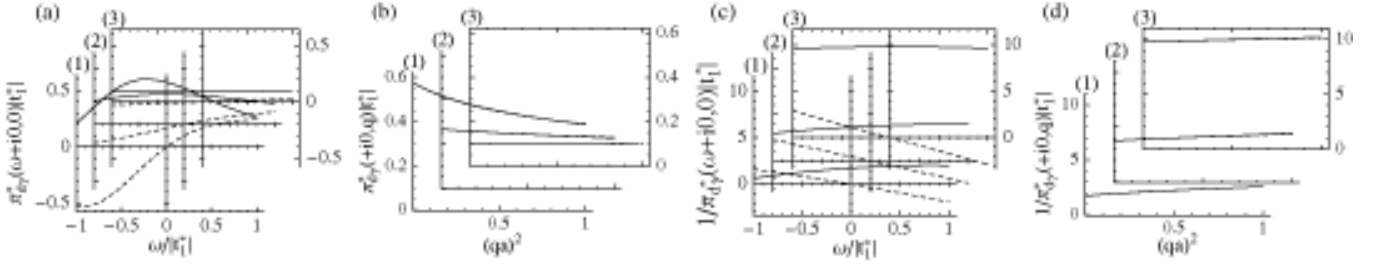} 
}
\caption[4]{ 
Polarization functions of $d\gamma$-wave Cooper-pair channel:
(a) $\pi_{d\gamma}^*(\omega+i0,0)|t_1^*|$,
(b) $\pi_{d\gamma}^*(+i0,{\bf q})|t_1^*|$,  
(c) $1/\pi_{d\gamma}^*(\omega+i0,0)|t_1^*|$ and
(d)  $1/\pi_{d\gamma}^*(+i0,{\bf q})|t_1^*|$ for
$n=0.9$ and $k_BT/|t_1^*|=0.1$. 
In Figs.~\ref{Fig_pi}(a)
and (c), solid and dotted lines show real and imaginary parts,
respectively. In Figs.~\ref{Fig_pi}(b)
and (d), solid and dotted lines show those for
${\bf q} \parallel (1,0)$ and  
${\bf q} \parallel (1,1)$, respectively; the polarization
functions are almost isotropic. In (1), (2) and (3), 
$\gamma/|t_1^*|=0.3$, 1 and 3 are assumed, respectively.
As $\gamma$'s increase, the
$\omega$-linear imaginary term of $1/\pi_{d\gamma}^*(\omega+i0,0)$
increases while its $q$-quadratic term decreases.
 }
\label{Fig_pi}
\end{figure*}  

According to Fermi-liquid relations\cite{Yamada,Yosida} and the
mapping condition (\ref{EqMapping}), it follows that 
$\bigl[\tilde{\chi}_s(0) \bigr]_{T \rightarrow +0} = 2 \tilde{W}_s
\rho_\gamma (0) $ in the limit of $\gamma\rightarrow 0$. In almost
half-filling cases $(n \simeq 1)$, $\tilde{W}_s \simeq 2$, as is
discussed in Sec.~\ref{prelim}.  According to
Fig.~\ref{rho_n_mu_FS}(a),  it follows that $\rho_\gamma (0) \simeq
0.2/|t_1^*|$ and  
$k_BT_K \simeq 1.2|t_1^*|$, unless the chemical potential is in the
vicinity of the two-dimensional van Hove singularity or $\gamma$'s
are extremely large. The specific heat coefficient due to local spin
and charge fluctuations is given by\cite{Yamada,Yosida}
$\tilde{\gamma}_C = (2/3)k_B^2 \tilde{\phi}_\gamma \rho_\gamma(0)$.

We take the following phenomenological expressions
for susceptibilities: 
\begin{subequations}\label{EqChiAFSC}
\begin{eqnarray}\label{EqChiAF}
&& \hspace*{-0.3cm} \chi_{s}(\omega+i0,{\bf Q}+{\bf q}) =
\frac{\chi_{s}(0,{\bf Q}) \kappa_{s}^2}
{\displaystyle \kappa_{s}^2 + (qa)^2 
-i \omega/\Gamma_s } ,
\\ \label{EqChidG2}  &&\hspace*{-0.3cm}
\chi_{d\gamma}^*(\omega+i0,{\bf q}) =
 \frac{\chi_{d\gamma}^*(0,0) \kappa_{d\gamma}^2}
{\displaystyle \kappa_{d\gamma}^2 + (qa)^2 
+ c_\omega \omega -i \omega/\Gamma_{d\gamma} } ,
\hspace{1.0cm}
\end{eqnarray}
\end{subequations}
with ${\bf Q} = (\pm\pi/a,\pm\pi/a)$.

According to an analysis of experimental data,\cite{Monthoux} 
$\chi_s(0,{\bf Q}) = 75~\mbox{states/eV}$,
$1/\kappa_s \simeq 2.3$ and
$\omega_{SF} = \Gamma_s \kappa_s^2 \simeq 14~\mbox{meV}$;
$\kappa_s = 0.43$, 
$\Gamma_s = 74~\mbox{meV}$ and 
$\chi_s(0,{\bf Q})\Gamma_s\kappa_s^2 = 1.1$. In the optimal-doped
region, the specific-heat coefficient is about 14~mJ/mol~K$^2$, so
that $|t_1^*| \simeq 50~\mbox{meV}$. Then, we assume
\begin{equation}\label{EqOptParaAF1}
\chi_{s}(0,{\bf Q})  \Gamma_s\kappa_{s}^2 \simeq 1, \quad 
\kappa_{s} \simeq \pi/8, \quad 
\Gamma_{s}/|t_1^*| \simeq  1, 
\end{equation}
for the optimal-doped region. 

When the dispersion relation of quasi-particles is given, it is
straightforward to calculate  polarization functions.
For example, Fig.~\ref{Fig_P} shows
$P(\omega+i0,{\bf q})$. Then, we can estimate some of parameters
appearing in Eq.~(\ref{EqChiAF}). According to 
Eqs.~(\ref{EqKondoSus2}) and (\ref{EqJQ2}), it follows that 
%
\begin{subequations}\label{PiS}
\begin{eqnarray}
&& \hspace*{-1.3cm}
\frac1{\chi_s(0,{\bf Q}) \Gamma_{s}
\kappa_{s}^2 } = - 
\frac {\tilde{W}_s^2}{\tilde{\chi}_s^2(0)} \mbox{Im}\!
\left[\frac{\partial P(\omega \!+\! i0,0)}{\partial \omega}
\right]_{\omega=0} \hspace{-0.0cm}, \label{PiS1} 
\\ && \hspace*{-1.3cm}
\frac1{\chi_{s}(0,{\bf Q}) \kappa_{s}^2 } 
= - \frac1{4}J +
\frac{\tilde{W}_s^2}{\tilde{\chi}_s^2(0)}
\left[\frac{\partial P(0,{\bf Q}\!+\!{\bf q})}{\partial (qa)^2}
\right]_{{\bf q}=0}  \hspace{-0.0cm}. \label{PiS2} 
\end{eqnarray}
\end{subequations}
Parameters estimated from Fig.~\ref{Fig_P} together with 
Eq.~(\ref{PiS}) for 
$\gamma/|t_1^*|=0.3$, $n=0.9$ and $k_BT/|t_1^*|=0.1$ are 
consistent with those given by Eq.~(\ref{EqOptParaAF1}). 

Fig.~\ref{Fig_pi} shows $\pi_{d\gamma}^*(\omega+i0,{\bf q})$.  
There exists an $\omega$-linear real term in 
$1/\pi_{d\gamma}^*(\omega+i0,{\bf q})$,  as is shown in
Fig.~\ref{Fig_pi}(c). We ignore it and we assume that
$c_\omega=0$ in Eq.~(\ref{EqChidG2}); nonzero $c_\omega$ can never
play any crucial role in physical properties examined in this paper.
According to Eqs.~(\ref{EqChiSC}) and (\ref{EqChidG2}), it follows
that
\begin{subequations}\label{PiDG}
\begin{eqnarray}
&& \hspace*{-1.2cm}
\frac{\tilde{W}_s^2}{\chi_{d\gamma}^*(0,0) \Gamma_{d\gamma}
\kappa_{d\gamma}^2 } = - \mbox{Im}\!
\left[\frac{\partial \phantom{\omega}}{\partial \omega}
\frac{1}{\pi_{d\gamma}^*(\omega \!+\! i0,0)}
\right]_{\omega=0} \hspace{-0.0cm}, \label{EqSCfluct1} 
\\ && \hspace*{-0.5cm}
\frac{\tilde{W}_s^2}{\chi_{d\gamma}^*(0,0) \kappa_{d\gamma}^2 } = 
\left[\frac{\partial \phantom{(qa)^2}}{\partial (qa)^2}
\frac{1}{\pi_{d\gamma}^*(0,{\bf q})}
\right]_{{\bf q}=0}  \hspace{-0.0cm}. \label{EqSCfluct2} 
\end{eqnarray}
\end{subequations}
Then, it follows from
Figs.~\ref{Fig_pi}(c) and (d) that
\begin{subequations}
\begin{equation}\label{EqOptParaSC1}
\chi_{d\gamma}^*(0,0) \Gamma_{d\gamma}
\kappa_{d\gamma}^2 \simeq \tilde{W}_s^2/2,  \quad 
\Gamma_{d\gamma}/|t_1^*| \simeq 1/2 ,
\end{equation}
for  $\gamma/|t_1^*|=0.3$, $n=0.9$ and $k_BT/|t_1^*| =0.1$; 
\begin{equation}\label{EqOptParaSC2}
\chi_{d\gamma}^*(0,0) \Gamma_{d\gamma}
\kappa_{d\gamma}^2 \simeq \tilde{W}_s^2/3,  \quad 
\Gamma_{d\gamma}/|t_1^*| \simeq 1/20 ,
\end{equation}
\end{subequations}
for $\gamma/|t_1^*|=3$, $n=0.9$ and $k_BT/|t_1^*| =0.1$. 

Note that there are crucial differences between effects
of $\gamma$ on spin and Cooper-pair channel
fluctuations: $J$ appears in Eq.~(\ref{PiS}) but it
does not in Eq.~(\ref{PiDG}); the
$\omega$-linear imaginary term becomes small with increasing $\gamma$
in the spin channel, as is shown in Fig.~\ref{Fig_P}(a), while it
does not in the Cooper-pair channel, as is
shown Fig.~\ref{Fig_pi}(c). For larger
$\gamma$'s,
$\Gamma_s$'s are larger but $\Gamma_{d\gamma}$'s are smaller. 
Small $\Gamma_{d\gamma}$'s  play a crucial role in
the formation of pseudo-gaps studied in Sec.~\ref{under-doped}.

When  we take Eq.~(\ref{EqChiAFSC}), the energy integration in
Eqs~(\ref{EqSelfAF}) and (\ref{EqSelfSC}) can be analytically
carried out: 
\begin{widetext} 
\begin{subequations}\label{EqSelf}
\begin{eqnarray}\label{EqSelfAF2}
\frac1{\tilde{\phi}_\gamma}
\Delta\Sigma_1^{(AF)}(\varepsilon+i0,{\bf k}) &=&
g_{AF} \frac1{N} \sum_{|{\bf q}|\le q_c} 
S\Bigl(\varepsilon, \xi({\bf k}+{\bf Q}+{\bf q}),
\Gamma_s \!\left[\kappa_s^2 +(qa)^2\right] \Bigr) ,
\\  \label{EqSelfSC20}
\frac1{\tilde{\phi}_\gamma}
\Delta\Sigma_1^{(SC)}(\varepsilon+i0,{\bf k}) &=&
g_{SC} \frac1{N} \sum_{|{\bf q}|\le q_c} 
\frac1{4}\eta_{d\gamma}^2 ({\bf k}
+\mbox{$\frac1{2}$}{\bf q}) 
S\Bigl(\varepsilon, -\xi({\bf k}+{\bf q}),
\Gamma_{d\gamma}\!\left[\kappa_{d\gamma}^2 +(qa)^2\right]\Bigr)
\\ \label{EqSelfSC2}  &\simeq&
g_{SC} \left[\frac1{4}\eta_{d\gamma}^2 ({\bf k})\right]
\frac1{N} \sum_{|{\bf q}|\le q_c}  
S\Bigl(\varepsilon, -\xi({\bf k}+{\bf q}),
\Gamma_{d\gamma}\!\left[\kappa_{d\gamma}^2 +(qa)^2\right]\Bigr) ,
\end{eqnarray}
\end{subequations}
with
\begin{equation}
g_{AF} = 3\left[\frac1{4}
I_s(\omega+i0,{\bf Q}) \tilde{W}_s \right]^2
\chi_s(0,{\bf Q}) \Gamma_s \kappa_s^2, \qquad
g_{SC}  = 
4 \left(\frac{3}{4}I_1^* \tilde{W}_s \right)^2
\chi_{d\gamma}^*(0,0) \Gamma_{d\gamma}\kappa_{d\gamma}^2, 
\end{equation}
%
%
\small
\begin{eqnarray}\label{EqS}
S(\varepsilon, x,\Gamma) &=&
- \frac{k_B T}{\Gamma}\frac1{\varepsilon+i\gamma-x+i\Gamma}
%
+ \frac1{2\pi} \left[
- \psi\! \left( \frac{\Gamma }{2\pi k_B T} \right)
+ \psi\! \left( 
\frac{ -i \varepsilon + i x + \gamma}{2\pi k_B T}\right) 
\right]
\frac1{\varepsilon + i \gamma - x - i \Gamma }
\nonumber \\ && 
+ \frac1{2\pi} \left[
\psi\! \left( 
\frac{-i\varepsilon \!+\! \Gamma }
{2\pi k_BT} \!+\! \frac1{2} \right) 
- \psi\! \left( \frac{\Gamma }{2\pi k_B T}
\right) 
- \psi\! \left( \frac{i x \!+\! \gamma}{2\pi k_B T}
\!+\! \frac1{2} \right) 
+ \psi\! \left(\frac{-i\varepsilon \!+\! i x \!+\! \gamma}
{2\pi k_BT}
 \right) \right]
\frac1{\varepsilon \!+\! i \gamma \!-\! x \!+\! i \Gamma } 
\nonumber \\ && 
+ \frac1{2\pi} \left[
- \psi\! \left( 
\frac{-i\varepsilon+\Gamma }{2\pi k_B T} + \frac1{2} \right)
+ \psi\! \left( \frac{ - i x + \gamma}{2\pi k_B T}
+ \frac1{2} \right) 
\right]
\frac1{\varepsilon  - i\gamma - x +i \Gamma} .
\end{eqnarray}
\normalsize
Because only low-energy antiferromagnetic fluctuations are crucial
in Eq.~(\ref{EqSelfAF2}),  we use instead of Eq.~(\ref{EqUAF})
\begin{equation}
U_{AF}(\omega \!+\! i0,{\bf q}) =
\frac{3}{4} \Bigl[ I_s(i\omega_l,{\bf q})-J({\bf q})
\Bigr] 
+ \frac{3}{4^2} I_s^2(i\omega_l,{\bf q})
\chi_s(i\omega_l,{\bf q}) 
\simeq 3 \left[\frac1{4} I_s(0,{\bf Q}) \right]^2 \!
\chi_s(\omega \!+\! i0,{\bf q}) .\hspace{1cm}
\end{equation}
\end{widetext}
Because we use the phenomenological forms, we restrict the
summation over {\bf q} in Eqs.~(\ref{EqSelf}) and we assume that 
\begin{equation}
q_ca = \pi/3.
\end{equation}
Because $q_c$ is rather small, we approximately use
Eq.~(\ref{EqSelfSC2}) instead of Eq.~(\ref{EqSelfSC20}).
In this approximation, there are two interesting properties:
When $\kappa_s=\kappa_{d\gamma}$ and
$\Gamma_s=\Gamma_{d\gamma}$, it follows that
\begin{subequations}
\begin{equation}\label{EqAF=SC}
\frac1{g_{AF}}\Delta\Sigma_\sigma^{(AF)}(+i0,{\bf k}_X)
=- \frac1{g_{SC}}\left[
\Delta\Sigma_\sigma^{(SC)}(+i0,{\bf k}_X)\right]^* \!,
\end{equation}
for $X$ points defined by ${\bf k}_X = (\pm\pi/a,\pm\pi/a)$.
In deriving Eq.~(\ref{EqAF=SC}), two relations of  
$S(0, x,\Gamma) = S^*(0, - x,\Gamma)$
and
$\xi(k_x+q_x,k_y+q_y) = \xi (
k_x+q_y \pm \pi/a, k_y+q_x \pm \pi/a )$
for ${\bf k}={\bf k}_X$ are made use of. The other is 
\begin{equation}\label{EqSCM}
\Delta\Sigma_\sigma^{(SC)}(\varepsilon+i0,{\bf k}) = 0,
\end{equation}
\end{subequations}
for ${\bf k} \parallel {\bf k}_M$, with ${\bf k}_M$ defined by
${\bf k}_M = (\pm\pi/a,0)$ and  $(0, \pm\pi/a)$. It is certain that
the renormalization by $d\gamma$-wave superconducting fluctuations is
anisotropic.

\begin{figure*}
\centerline{
\includegraphics[width=16cm]{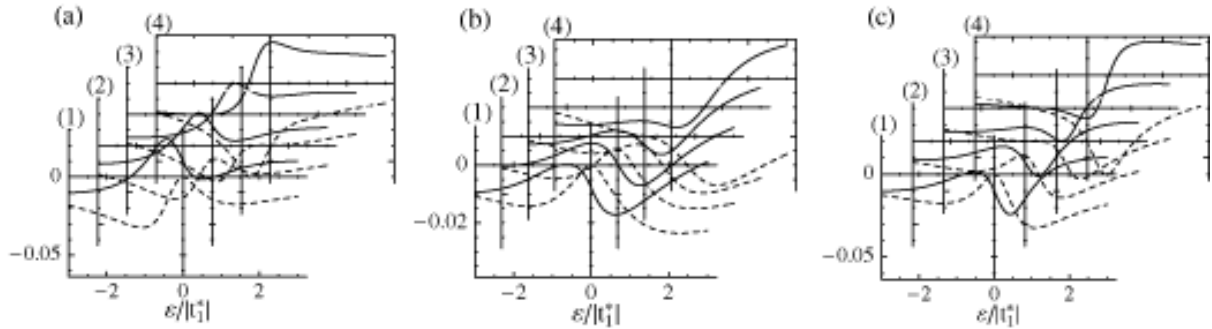}
}
\caption[5]{ 
Self-energy corrections:
(a) $\Sigma_\sigma^{(AF)}(i\varepsilon+i0,{\bf k}_X)
|t_1^*|/\tilde{\phi}_\gamma g_{AF}$ and  
(b) $\Sigma_\sigma^{(AF)}(i\varepsilon+i0,{\bf k}_M/2)
|t_1^*|/\tilde{\phi}_\gamma g_{AF}$ 
for $n=0.9$, $\kappa_s=\pi/8$ and $\Gamma_s/|t_1^*|=1$;  
(c) $\Sigma_\sigma^{(SC)}(i\varepsilon+i0,{\bf k}_X) 
|t_1^*|/\tilde{\phi}_\gamma g_{SC}$
for $n=0.9$, $\kappa_{d\gamma}=\pi/8$ and
$\Gamma_{d\gamma}/|t_1^*|=1$. In (1), (2), (3) and (4), 
$k_BT/|t_1^*|=0.1$, 0.3, 1 and 3 are assumed, respectively;
solid and dotted lines show the real and imaginary parts. }
\label{Fig_self1}
\end{figure*}   
\begin{figure*} 
\centerline{
\includegraphics[width=16cm]{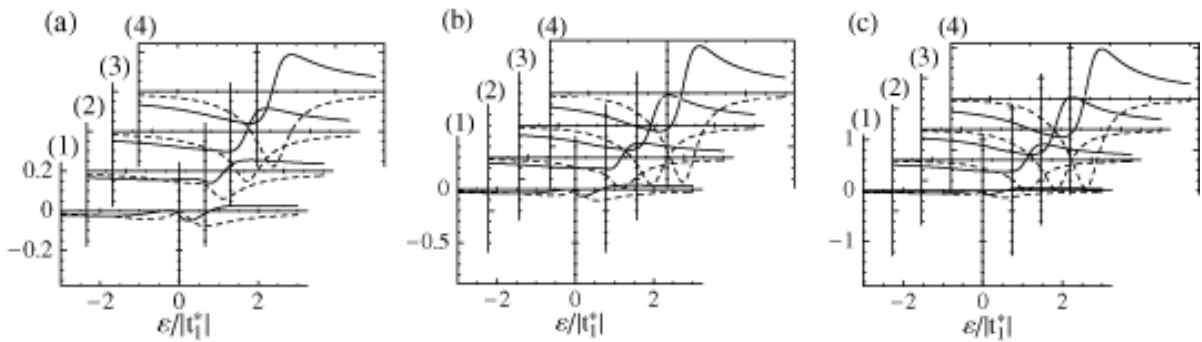}
}
\caption[6]{ 
Self-energy corrections:
$\Sigma_\sigma^{(SC)}(i\varepsilon+i0,§{\bf k}_X)
|t_1^*|/\tilde{\phi}_\gamma g_{SC}$ 
for $n=0.9$, $\kappa_{d\gamma} = \pi/8$ and smaller
$\Gamma_{d\gamma}$'s;  
(a) $\Gamma_{d\gamma}/|t_1^*|=1/4$,
(b) $\Gamma_{d\gamma}/|t_1^*|=1/8$, and
(c) $\Gamma_{d\gamma}/|t_1^*|=1/16$.
In (1), (2), (3) and (4), 
$k_BT/|t_1^*|=0.1$, 0.3, 1 and 3 are assumed, respectively;
solid and dotted lines show the real and imaginary parts. 
}
\label{Fig_self2}
\end{figure*} 

\begin{figure*} 
\centerline{\includegraphics[width=17.8cm]{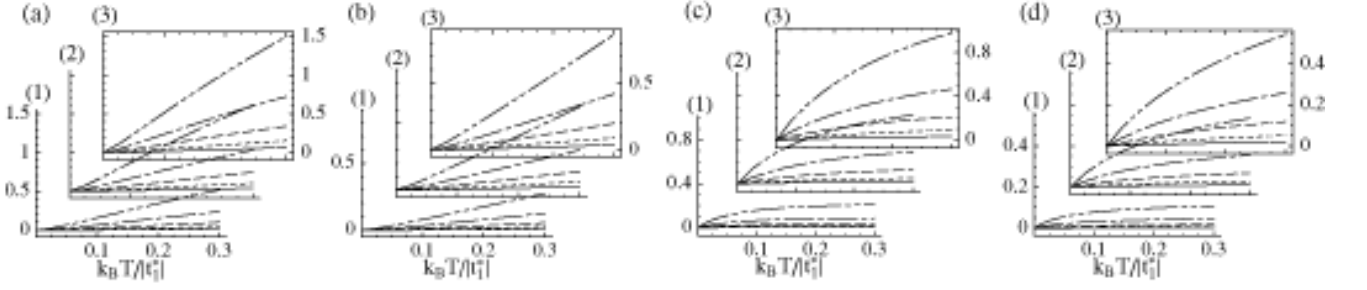}}
\caption[7]{ 
Inelastic life-time widths:
 Figs.~\ref{Fig_selfT}(a) and (c) show
$\mbox{Im} \bigl[ \Delta\Sigma_1^{(AF)}(+i0,{\bf k}_X)
\bigr]|t_1^*|/\tilde{\phi}_\gamma g_{AF}$, which is equal to
$-\mbox{Im} \bigl[ \Delta\Sigma^{(SC)}(+i0,{\bf k}_X)
\bigr]^*|t_1^*|/\tilde{\phi}_\gamma g_{SC}$ when their
corresponding parameters are the same between the two channels;  
 Figs.~\ref{Fig_selfT}(b) and (d) show
$\mbox{Im} \bigl[ \Delta\Sigma_1^{(AF)}(+i0,{\bf k}_M/2)
\bigr]|t_1^*|/\tilde{\phi}_\gamma g_{AF}$. 
In each figure, $n=0.9$ and $\gamma/|t_1^*|=0.3$ are assumed;
solid,  dotted, broken, dot-broken and two-dot-broken lines are for
$\Gamma_s/|t_1^*|=1$, 1/2, 1/4, 1/8 and 1/16, respectively.
In Figs.~\ref{Fig_selfT}(a) and (b), $\kappa_s$ is assumed to be
independent of $T$:  
(1) $\kappa_s=\pi/4$, 
(2) $\kappa_s=\pi/8$, and
(3) $\kappa_s=\pi/12$.
In Figs.~\ref{Fig_selfT}~(c) and (d), $\kappa_s$ is given by
Eq.~(\ref{EqT-depKap});  (1) $\kappa_0=\pi/4$, (2)
$\kappa_0=\pi/8$, and (3) $\kappa_0=\pi/12$. }
\label{Fig_selfT}
\end{figure*}  
\begin{figure}
\centerline{\includegraphics[width=6.5cm]{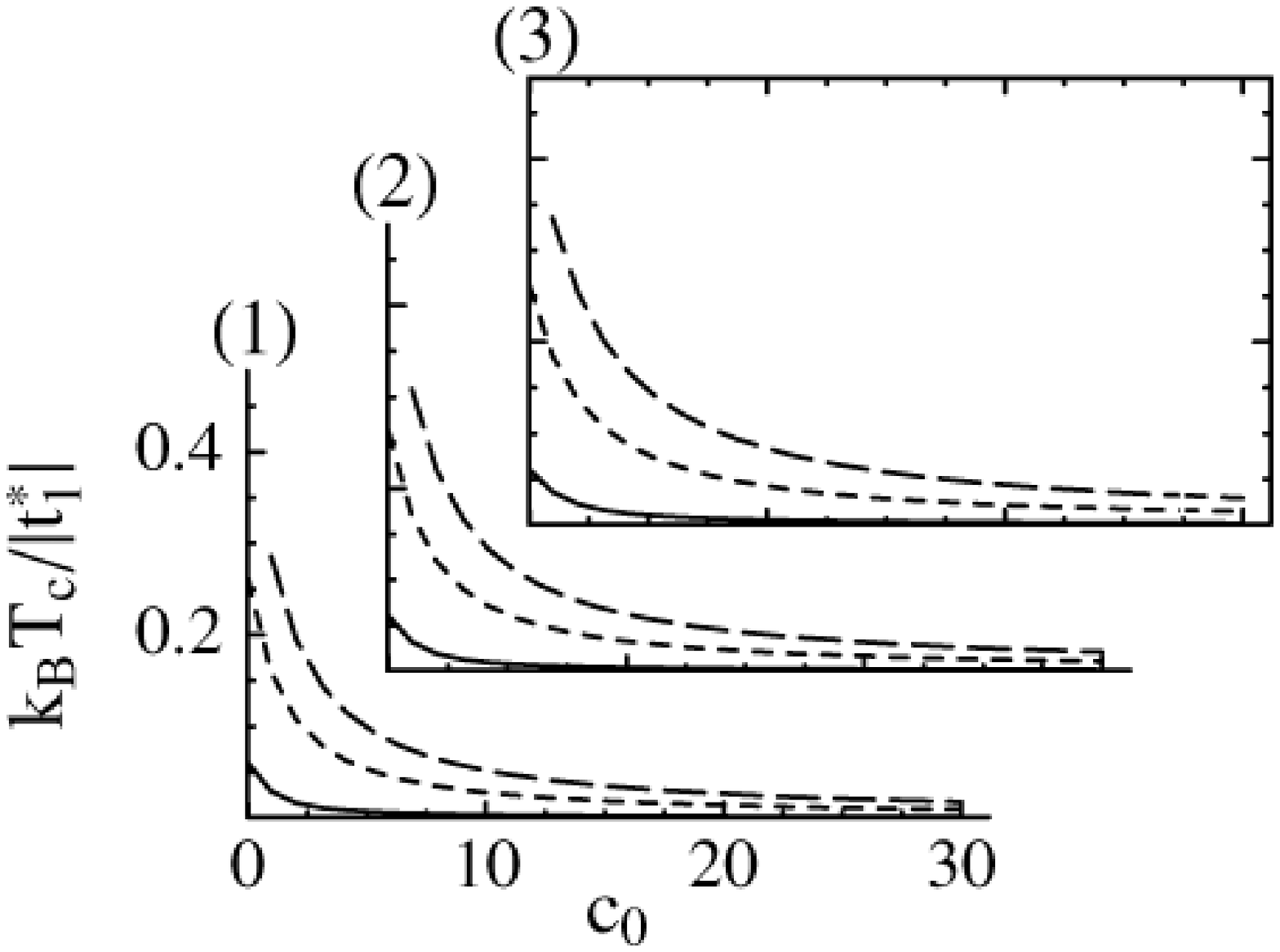}}
\caption[8]{ 
Superconducting $T_c$ as a function of the $T$-linear coefficient
$c_0$: (1) $\gamma = c_0 k_BT$, (2) $\gamma = c_0 k_B
T \left[1 + \eta_{d\gamma}^2({\bf k})/4 \right]/2$, and (3)
$\gamma = c_0 k_B T \eta_{d\gamma}^2({\bf k})/4$. 
 Solid, dotted, and broken lines show $T_c$ for
$I_1^* \tilde{W}_s^2/|t_1^*| =-1, -2$, and $-3$, respectively. }
\label{Fig_Tc}
\end{figure} 

The electron density is crucial for the nesting of the Fermi surface
and the spin susceptibility. Because we use the phenomenological
form (\ref{EqChiAF}) for the spin susceptibility, however, the
electron density itself is never crucial. Then, we assume $n=0.9$
here and in the following part of this paper. 
Fig.~\ref{Fig_self1}(a) shows
$\Delta\Sigma_\sigma^{(AF)}(\varepsilon\!+\!i0,{\bf k}_X)$, 
Fig.~\ref{Fig_self1}(b) shows
$\Delta\Sigma_\sigma^{(AF)}(\varepsilon\!+\!i0,{\bf k}_M/2)$, and
Fig.~\ref{Fig_self1}(c) and Fig.~\ref{Fig_self2} show
$\Delta\Sigma_\sigma^{(SC)}(\varepsilon\!+\!i0,{\bf k}_X)$.
Fig.~\ref{Fig_selfT} shows the imaginary parts of their static
components $( \varepsilon=+i0)$ or inelastic life-time widths of
quasi-particles.

Because $|t_1^*| \simeq 50~\mbox{meV}$,
\begin{equation}
|t_1/t_1^*| \simeq 10
\end{equation}
in the optimal-doped region.
If only the  superexchange interaction as large as
Eq.~(\ref{EqJ}) is considered, it follows that
$|I_s(0,{\bf Q}|)/4|t_1^*| > |J|/|t_1^*| = (2\mbox{-}3)$ and 
$|I_1^*|/|t_1^*| > |J|/|t_1^*| =(2\mbox{-} 3)$.
When we take  Eqs.~(\ref{EqOptParaAF1}) and (\ref{EqOptParaSC1}),
$g_{AF}= (12 \mbox{-} 27)\tilde{W}_s^2|t_1^*|^2$ and
$g_{SC} = (4.5 \mbox{-} 10)\tilde{W}_s^4|t_1^*|^2$.  If we
take $\tilde{W}_s \simeq 2$, 
$g_{AF} \simeq (48\mbox{-}100) |t_1^*|^2$ and  
$g_{SC} \simeq (72\mbox{-}160) |t_1^*|^2$.
On the other hand, 
observed temperature dependence of resistivity implies 
\begin{equation}\label{EqSTdepObs}
(1/\tilde{\phi}_\gamma)
\mbox{Im}\ \Delta\Sigma_\sigma(+i0,{\bf k})
\simeq -(1\mbox{-}2) k_BT ,
\end{equation}
in the optimal-doped
region, so that $g_{AF}$ and $g_{SC}$ should be
much smaller than these.
For example, if we take
\begin{equation}\label{EqWs}
\tilde{W}_s \simeq (0.7\mbox{-} 1) ,
\end{equation}
$g_{AF} \simeq 8 (t_1^*)^2$ and 
$g_{SC} \simeq 2 (t_1^*)^2$.
Then, it follows from Figs.~\ref{Fig_selfT}(a) and (b) that
\begin{subequations}\label{EqSTdep}
\begin{equation}\label{EqSTdepAF}
(1/\tilde{\phi}_\gamma)
\mbox{Im}\ \Delta\Sigma_\sigma^{(AF)}(+i0,{\bf k})
\simeq - k_BT ,
\end{equation}
for ${\bf k}={\bf k}_X$ and ${\bf k}={\bf k}_M/2$;
it follows from  Fig.~\ref{Fig_selfT}(a) that
\begin{equation}\label{EqSTdepSC}
(1/\tilde{\phi}_\gamma)
\mbox{Im}\ \Delta\Sigma_\sigma^{(SC)}(+i0,{\bf k}_X)
\simeq - k_BT .
\end{equation}
\end{subequations}
Eq.~(\ref{EqSTdep}) is consistent with
Eq.~(\ref{EqSTdepObs}). As long as $\tilde{W}_s
\simeq 2$, we cannot explain observed resistivity of the
optimal-doping region. 

There is a similar drawback for theoretical superconducting
critical temperatures $T_c$.
According to Eq.~(\ref{EqChiSC}), $T_c$'s are given by
\begin{equation}\label{EqTcSC}
1 + \frac{3}{4}I_1^*\tilde{W}_s^2 \pi^*(0,0) =0.
\end{equation}
The imaginary part of the self-energy has a reduction effect of
$T_c$. Because its static part plays the most crucial role, we
consider it as the pair breaking. Because it is almost
linear in $T$ as is shown in Fig.~\ref{Fig_selfT}, we assume a
phenomenological form:
\begin{equation}\label{EqPairBreak}
\gamma = c_{0} \left[ c_1 + (1-c_1)\frac1{4}\eta_{d\gamma}^2
({\bf k}) \right] k_B T .
\end{equation}
 Fig.~\ref{Fig_Tc} shows  $T_c$ for three
cases of $c_1=0$, $1/2$ and $1$  as a function of $c_0$. The
reduction of $T_c$ is as large as 
$T_{c0}/T_c \simeq 2$ for $c_0\simeq (1$-2),
with
$T_{c0}$ critical temperatures in the absence of any pair
breaking.  If we take 
\begin{equation}\label{EqOptPara2}
I_1^*\tilde{W}_s^2/|t_1^*| = -(1.5 \mbox{-} 3), 
\end{equation}
observed $T_c$ in the
optimal-doped region can be explained.  
Eq.~(\ref{EqOptPara2}) is satisfied 
when we take Eq.~(\ref{EqWs}). 

In the formulation in Sec.~\ref{prelim}, 
$g_\sigma^{(0)}(i\varepsilon_n,{\bf k})$ appear in Feynman diagrams.
These 
$g_\sigma^{(0)}(i\varepsilon_n,{\bf k})$ had better be renormalized
into $g_\sigma(i\varepsilon_n,{\bf k})$ in an improved theory. 
When this renormalization is made,\cite{OhkawaSC2}  
$\tilde{\phi}_\gamma$ is renormalized into 
$\phi_\gamma({\bf k})$ and $\tilde{W}_s$ into
$W_s({\bf k})  \equiv \tilde{\phi}_s/\phi_\gamma({\bf k})$. The
energy derivative of the intersite self-energy is estimated from
results shown in Figs.~\ref{Fig_self1} and \ref{Fig_self2}, so
that  $W_s({\bf k})$ must be significantly smaller than
$\tilde{W}_s \simeq 2$.\cite{vertex}  
In this paper, however, we simply treat $\tilde{W}_s$ as another
phenomenological parameter; we assume
Eq.~(\ref{EqWs}), that is, 
$\tilde{W}_s \simeq (0.7\mbox{-}1)$.

\subsection{Pseudo-gaps and {\em low}-$T_c$ superconductivity}
\label{under-doped}

The following four properties  are crucial for the under-doped region. 
(i) Observed superconducting gaps
$\epsilon_G(0)$ at $T=0$~K increase with decreasing {\em hole}
dopings.\cite{Shen2,Shen3,Ding,Ino,Renner,Ido1,Ido2,Ido4,Ekino} It implies
an increase of the Cooper-pair interaction, that is, the enhanced exchange
interaction $I_1^*$.  It is also theoretically argued in
Sec.~\ref{discussion} that $I_1^*$ increases with decreasing {\em hole}
dopings.   (ii) As is discussed in Sec.~\ref{high-Tc}, the energy scale of
superconducting fluctuations $\Gamma_{d\gamma}$ is smaller for a larger
life-time width $\gamma$ of quasi-particles.  (iii) The life-time width
$\gamma$ is larger for smaller $\Gamma_{d\gamma}$ or when low-energy
superconducting fluctuations are developed, as is shown in
Fig.~\ref{Fig_selfT}.  (iv) As {\it hole} dopings decrease, cuprate oxides
are closer to an antiferromagnetic phase and inelastic scatterings by
antiferromagnetic fluctuations more substantially contribute to
$\gamma$.  Then, we claim that the under-doped region can be characterized
by large $\gamma$ and small $\Gamma_{d\gamma}$, or large life-time widths
of quasi-particles and low-energy superconducting fluctuations.  

An increase
of $\gamma$ causes a decrease of $\Gamma_{d\gamma}$ and the decrease of
$\Gamma_{d\gamma}$ causes an increase of $\gamma$.  Because of this
cooperative effect, an increase of $I_1^*$ and a development of
antiferromagnetic fluctuations with decreasing {\em hole} dopings causes
an increase of $\gamma$ and it eventually causes a large decrease of
$\Gamma_{d\gamma}$ and a large increase of $\gamma$.  Although this
cooperative effect should be examined in a self-consistent way, we treat
$\Gamma_{d\gamma}$ as another phenomenological parameter; we assume
that it is as small as 
\begin{equation}
1/64 \alt \Gamma_{d\gamma}/|t_1^*|
\alt 1/2. 
\end{equation}
in the under-doped region.

The life-time width $\gamma$ makes the $\omega$ and {\bf q}  dependences
of $P(\omega+i0,{\bf q})$ small, as is shown in Figs.~\ref{Fig_P}(a)
and (b).  Then, the $\omega$ dependence of 
$1/\chi_s(\omega+i0,{\bf q})$ is small in the under-doped region.
However, its {\bf q} dependence  can never be as small as that of 
$1/\chi_{d\gamma}^*(\omega+i0,{\bf q})$,  because the
${\bf q}$ dependence of the superexchange interaction is never
suppressed by $\gamma$. Therefore, $\Gamma_s$'s must
be large rather than small in the under-doped region.  However, we
assume for the sake of simplicity that $\Gamma_s/|t_1^*| = 1$ even
in the under-doped region.

Although $g_{AF}$ and $g_{SC}$ for the under-doped region  are
larger than those for the optimal-doped one, we assume similar
values to those for the optimal-doped one:
\begin{equation}\label{EqOptPara}
g_{AF} = 8 (t_1^*)^2 , \quad
g_{SC} = 2 (t_1^*)^2 .
\end{equation}

\begin{figure*}
\centerline{\hspace*{0.5cm}
\includegraphics[width=18.0cm]{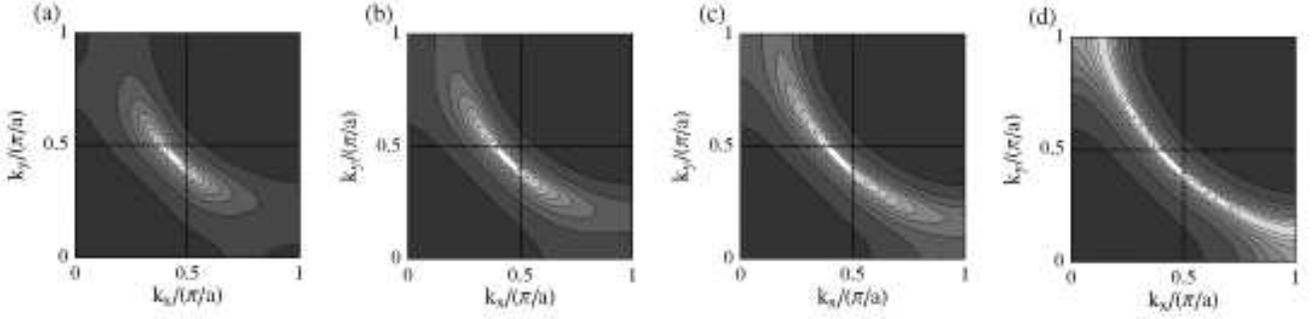}
}
\caption[9]{ 
Relative intensities $\rho_{\bf k} (0)$ as functions of $k_x$ and
$k_y$ for $n=0.9$, $k_BT/|t_1^*|=0.1$,
$\kappa_s=\kappa_{d\gamma}=(\pi/8)$, 
$\Gamma_s/|t_1^*|=1$, $g_{AF}=8|t_1^*|^2$ and $g_{SC}=2|t_1^*|^2$:
(a) $\Gamma_{d\gamma}/|t_1^*|=1/64$,
(b) $\Gamma_{d\gamma}/|t_1^*|=1/32$,
(c) $\Gamma_{d\gamma}/|t_1^*|=1/16$, and 
(d) $\Gamma_{d\gamma}/|t_1^*|=1/2$.
 When
$\Gamma_{d\gamma}$ are small, quasi-particles are well defined only
around  ${\bf k}=(\pi/2a,\pi/2a)$. Lighter parts have large
intensities. The well-defined
region extends toward to $(\pi/a,0)$ and $(0,\pi/a)$ as
$\Gamma_{d\gamma}$ becomes larger. }
\label{FIg_HC.EPS}
\end{figure*} 
\begin{figure} 
\centerline{\hspace*{0.5cm}
\includegraphics[width=9.5cm]{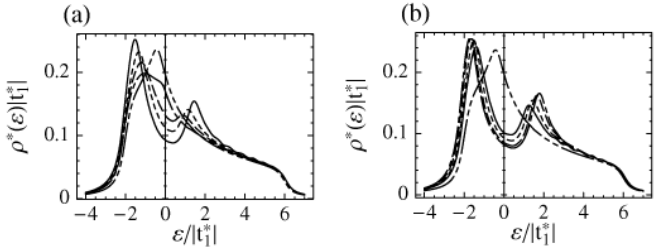}
}
\caption[10]{ 
Density of states $\rho^*(\varepsilon)$. 
In Fig.~\ref{Fig_gap}(a), parameters
are the same as those for Fig.~\ref{FIg_HC.EPS}:
solid,  dotted, broken and dot-broken lines correspond to
Fig.~\ref{FIg_HC.EPS}(a), (b), (c) and (d), respectively.
In Fig.~\ref{Fig_gap}(b), we assume that
$\Gamma_s/|t_1^*|=1$, $\Gamma_{d\gamma}/|t_1^*|=1/64$ and
$\kappa_s = \kappa_{d\gamma}=(\pi/8) \sqrt{10 k_BT/|t_1^*|}$; 
solid,  dotted, broken and dot-broken lines are
for $k_BT/|t_1^*|=0.3$, $0.2$, $0.1$ and $0.05$, respectively.  In
both of Figs.~\ref{Fig_gap}(a) and (b), $\rho_\gamma(\varepsilon)$
for
$\gamma/|t_1^*|=0.3$ is shown in a two-dot-broken line for
comparison.  In Fig.~\ref{Fig_gap}(a), pseudo-gaps become larger
with decreasing $\Gamma_{d\gamma}$. In Fig.~\ref{Fig_gap}(b), they
become smaller with decreasing $T$.
}
\label{Fig_gap}
\end{figure} 
\begin{figure*}
\includegraphics[width=18cm]{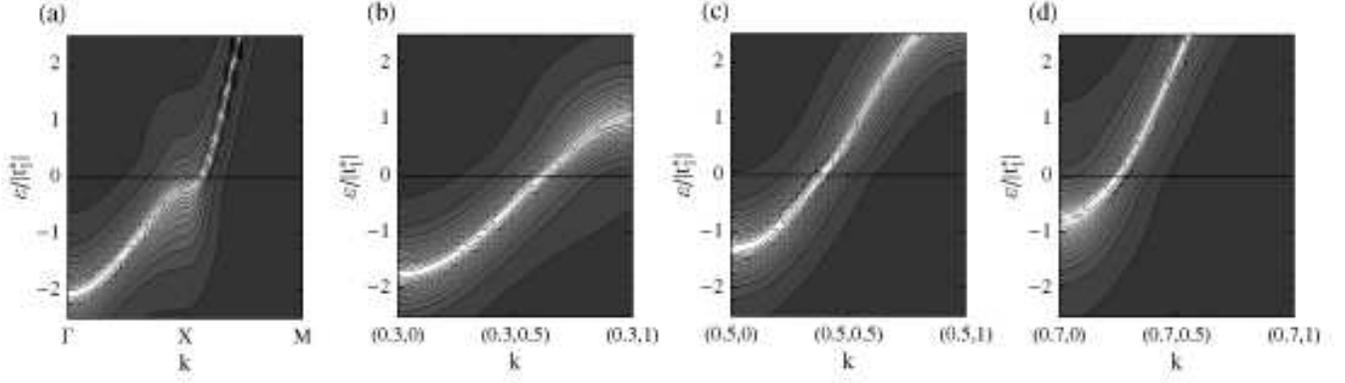}
\caption[11]{ 
Relative intensities of  $\rho_{\bf k}(\varepsilon)$.
Parameters are the same as those for
Fig.~\ref{FIg_HC.EPS}(d), and they correspond to those for the
optimal-doped region except for the electron density $n=0.9$.   
Lighter parts have large intensities. 
}
\label{Fig_peak1}
\end{figure*}
\begin{figure*}
\centerline{\includegraphics[width=18cm]{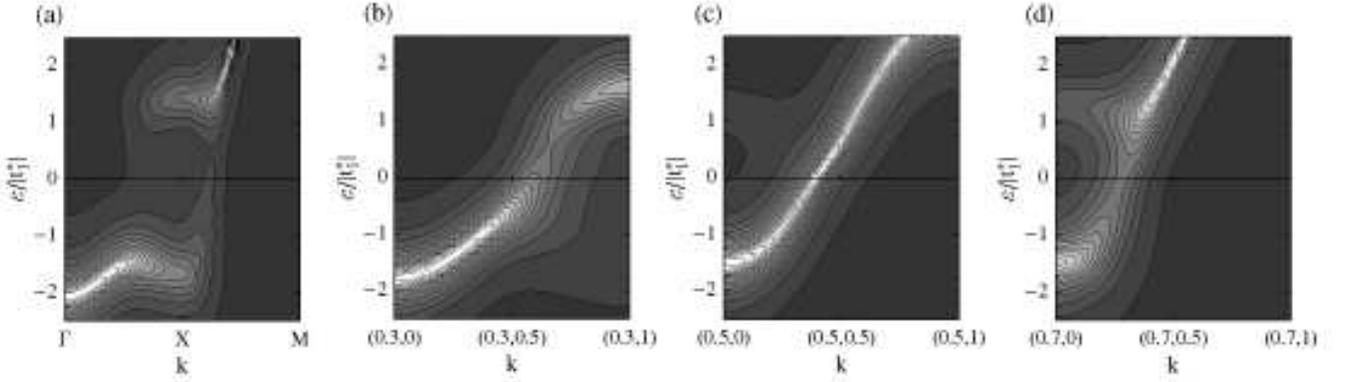}}
\caption[12]{ 
Relative intensities of  $\rho_{\bf k}(\varepsilon)$. Parameters are
the same as those for Fig.~\ref{FIg_HC.EPS}(a), and they correspond
to those for the under-doped doping region;   
the electron density is assumed to be $n=0.9$.
Lighter parts have large intensities.
}
\label{Fig_peak2}
\end{figure*}

Figs.~\ref{FIg_HC.EPS}(a)-(d) show relative intensities of
quasi-particles at the chemical potential,
$\rho_{\bf k}^*(0)$, in a fourth of the
Brillouin zone for four cases such as 
(a) $\Gamma_{d\gamma}/|t_1^*| =1/64$, 
(b) $\Gamma_{d\gamma}/|t_1^*| =1/32$, 
(c) $\Gamma_{d\gamma}/|t_1^*| =1/16$, and
(d) $\Gamma_{d\gamma}/|t_1^*| =1/2$; 
other parameter are $\gamma =0.3|t_1^*|$, $k_BT/|t_1^*|=0.1$ and
$\kappa_s =\kappa_{d\gamma} =\pi/8$.  When $\Gamma_{d\gamma}$ are
small, spectral intensities are swept away and are almost vanishing 
around $X$ points, $(\pm\pi/a,0)$ and
$(0,\pm\pi/a)$,
as is shown in Fig.~\ref{FIg_HC.EPS}(a). The vanishing of
quasi-particle spectra around $X$ points is because of large
life-time widths. When $\Gamma_{d\gamma}$ are
large, on the other hand, life-time widths are small and the
Fermi surface or the Fermi line forms a closed line, as is shown in
Fig.~\ref{FIg_HC.EPS}(d).

Fig.~\ref{Fig_gap}(a) shows the density of
states for quasi-particles, $\rho^*(\varepsilon)$, together with
$\rho_{\gamma=0.3|t_1^*|}(\varepsilon)$ for comparison, as a
function of 
$\varepsilon$ for four values of $\Gamma_{d\gamma}$:
$\Gamma_{d\gamma}/|t_1^*| =1/64$, $1/32$, $1/16$, and $1/2$; other
parameter are the same as those for  Fig.~\ref{FIg_HC.EPS}.   
Pseudo-gap develop as $\Gamma_{d\gamma}$ decrease.

Figs.~\ref{Fig_peak1} and \ref{Fig_peak2} show relative intensities
of $\rho_{\bf k}^*(\varepsilon)$ as a function of
$\varepsilon$ and ${\bf k}$. Parameters used  for
Figs.~\ref{Fig_peak1} and \ref{Fig_peak2} are the same as those used
for Figs.~\ref{FIg_HC.EPS}(d) and (a), respectively;
Fig.~\ref{Fig_peak1} presumably corresponds to the optimal-doped
region and  Fig.~\ref{Fig_peak2} to the under-doped region. A
weakly dispersive band around $X$ points in Fig.~\ref{Fig_peak1}(a)
is because of saddle points in the dispersion relation of
quasi-particles. The spectra shown in Figs.~\ref{Fig_peak1}(a)-(d)
are rather normal because life-time widths are small. On the
other hand, we can see interesting and anomalous properties in 
Figs.~\ref{Fig_peak2}(a)-(d) because life-time widths
large. Large pseudo-gaps open around
$X$ points, and almost dispersionless peaks or {\it flat} bands
appear along $\Gamma$-$X$ below and above the chemical potential,
as is shown in Fig.~\ref{Fig_peak2}(a). Such flat
bands much below the chemical potential must correspond to 
observed flat bands.\cite{Ino} Pseudo-gaps become smaller as
wavenumbers are closer to ${\bf k}_M/2$, as is shown in
Figs.~\ref{Fig_peak2}(b)-(d).  

\section{Discussions}
\label{discussion}

In an improved theory, the effective transfer integral $t_1^*$ of
quasi-particles is further renormalized by antiferromagnetic and
superconducting fluctuations.  
In this paper, however, we treat $t_1^*$ for the {\em unperturbed}
state as one of phenomenological parameters and we take it as the
energy unit. Therefore, $T$-dependences discussed in this paper are
qualitative. 

{\em Low}-$T_c$ superconductivity in the under-doped region can be
explained by large inelastic life-time widths of quasi-particles.
For example, 
$(1/\tilde{\phi}_\gamma)\mbox{Im}\ \Delta\Sigma_\sigma(+i0, {\bf
k}_X) \simeq -(10\mbox{-}20) k_BT$ for parameters used to obtain
Fig.~\ref{Fig_peak2}. As is shown in Fig.~\ref{Fig_Tc}, the
reduction of $T_c$ is very large for such large life-time widths.  

When superconducting gaps as large as $\epsilon_{G}(T)$ open, spin
and superconducting low-energy fluctuations with
$|\omega|\alt\epsilon_{G}(T)$ are depressed and the pair breaking
caused by the fluctuations must also be depressed. Therefore, the
development of $\epsilon_{G}(T)$ must be very rapid with
decreasing $T$. This argument is  consistent with
experiment.\cite{Ekino,Ido4} It also implies that the reduction of
$\epsilon_G(0)$ by inelastic scatterings must be very small, so that
\begin{equation}
\frac{\epsilon_G(0)}{k_BT_c}  = \frac{\epsilon_G(0)}{k_BT_{c0}} 
\frac{T_{c0}}{T_{c}} \simeq 4 \frac{T_{c0}}{T_{c}},
\end{equation}
with $T_{c0}$ critical temperatures in the absence of any
pair breaking. In Sec.~\ref{high-Tc}, we argue that $T_{c0}/T_c
\simeq 2$ in the optimal region, so that $\epsilon_G(0)/k_BT_c
\simeq 8$. This number 8 is consistent with experiment.\cite{Ido4}
It is quite reasonable that $\epsilon_G(0)/k_BT_c > 8$ or 
$\epsilon_G(0)/k_BT_c \gg 8$ in the under-doped region.

It is likely that $\kappa_s^2$ and $\kappa_{d\gamma}^2$ decrease 
almost linearly in $T$ with decreasing $T$ and they vanish at
critical temperatures, $T_N$ and $T_c$, for antiferromagnetism and
superconductivity, respectively. Then, we assume that both of $T_N$
and $T_c$ are zero or very low, so that 
\begin{equation}\label{EqT-depKap}
\kappa_s = \kappa_0 \sqrt{10 k_BT/|t_1^*|}, \quad
\kappa_{d\gamma} = \kappa_0 \sqrt{10 k_BT/|t_1^*|}. 
\end{equation}
We ignore the $T$-dependence of $\Gamma_s$ and $\Gamma_{d\gamma}$.
Figs.~\ref{Fig_selfT}(c) and (d) show inelastic life-time
widths of quasi-particles for three cases of $\kappa_0=\pi/4$,
$\pi/8$ and $\pi/12$. It decreases sublinearly  in
$T$ rather than linearly. It is likely that resistivity varies in
$T^\alpha$ with $\alpha<1$ in the under-doped region.  
It is desirable to develop a microscopic theory in order to
predict a precise number of the exponent.

One may argue that the opening of pseudo-gaps is evidence for the
formation of preformed Cooper pairs above $T_c$. If this scenario is
the case, pseudo-gaps must increase with decreasing $T$ at any
temperature region. Experimentally, however, pseudo-gaps decrease
with decreasing $T$ at a temperature range close to 
$T_c$.\cite{Ekino,Ido4}  This observation contradicts the above
scenario. Instead, it is a piece of evidence that the opening of
pseudo-gaps is because spectral intensities of quasi-particles around
$X$ points are swept away by inelastic scatterings.
Fig.~\ref{Fig_gap}(b) shows
$\rho^*(\varepsilon)$,  together with
$\rho_{\gamma=0.3|t_1^*|}(\varepsilon)$ for comparison, as a
function of 
$\varepsilon$ for four cases such as
$k_BT/|t_1^*| =0.3$, $0.2$, $0.1$, and $0.05$;
we also assume Eq.~(\ref{EqT-depKap}) with $\kappa_0=\pi/8$ as well
as $\Gamma_s/|t_1^*|=1$ and  $\Gamma_{d\gamma}/|t_1^*|=1/64$. The
decrease of $\kappa_{d\gamma}$ causes an increase of pseudo-gaps,
if other parameters are constant. In Fig.~\ref{Fig_gap}(b), however,
the magnitude of pseudo-gaps decreases with decreasing $T$ because
the decrease of life-time widths  dominates that of
$\kappa_{d\gamma}$. This result is consistent with
experiment.\cite{Ekino,Ido4}  

It is reasonable that $\kappa_{d\gamma}^2$ is much smaller below
$T_{c0}$ than it is above 
$T_{c0}\simeq\epsilon_{G}(0)/(4k_B)$ or that the coefficient of the
linear $T$ term in  $\kappa_{d\gamma}^2$ is quite different between
below and above $T_{c0}$. If this is actually the case,  pseudo-gaps
start to open around $T_{c0}$. It is desirable to study the
$T$-dependences of $\Gamma_{d\gamma}$ and $\kappa_{d\gamma}$  in
order to predict the $T$-dependence of pseudo-gaps.

When Eq.~(\ref{EqChiAF}) is assumed,
the longitudinal NMR rate is given by
\begin{equation}
\frac1{T_1T} \propto \frac1{N}\sum_{\bf q}
\lim_{\omega\rightarrow+0}
\mbox{Im} \frac{\chi_s(\omega \!+\! i0,{\bf q})}{\omega} 
= \frac{\chi_s(0,{\bf Q})}{\Gamma_s}.
\end{equation}
Because life-time widths are large in the under-doped region,
$\Gamma_s$'s are also large there. When the opening of pseudo-gaps
is mainly caused by superconducting fluctuations, $\chi_s(0,{\bf Q})$ must be
reduced and $\Gamma_s$ must be enhanced. It is interesting to
examine if the development of superconducting fluctuations can
actually explain observed reduction of $1/T_1T$.\cite{yasuoka} It is
difficult to explain the reduction of $\chi_s(0,{\bf q})$ and
$1/T_1T$ by the development of antiferromagnetic spin fluctuations.

Experimentally, the so called kink structure is observed in the
dispersion relation of quasi-particles determined by ARPES
experiment.\cite{lanzara} Very tiny kink structures can be seen
around the chemical potential in Figs.~\ref{Fig_peak1}(c) and
\ref{Fig_peak2}(c); they are too tiny to explain ARPES experiment.
When pseudo-gaps or superconducting gaps are developed, spectra of
antiferromagnetic and superconducting fluctuations may have certain
gap-like structures, which are often called resonance modes. It is
interesting to examine if such resonance modes actually exist in
$\chi_{s}(\omega+i0,{\bf q})$  and
$\chi_{d\gamma}^*(\omega+i0,{\bf q})$. One may argue that what cause
pseudo-gap must cause the kink structure. It is interesting to
examine effects of  superconducting fluctuations
extending in the wavenumber space, whose energies are rather high, as
large as
$k_BT_K$ or of the order of the bandwidth of quasi-particles. 

\begin{figure} 
\centerline{\hspace*{0.1cm}
\includegraphics[width=8.3cm]{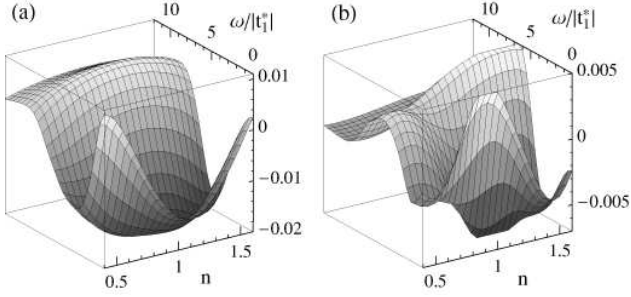}
}
\caption[13]{ 
(a) Nearest neighbor  
$\mbox{Re}\bigl[|t_1^*| P_1(\omega+i0) \bigr]$ and 
(b) next-nearest neighbor 
$\mbox{Re}\bigl[|t_1^*| P_2(\omega+i0) \bigr]$
as a function of $\omega$ and $n$ for
$k_BT/|t_1^*|=0.1$ and $\gamma/|t_1^*|=1$. 
  }
\label{Fig_exch1}
\end{figure} 
\begin{figure}
\centerline{
\includegraphics[width=8.0cm]{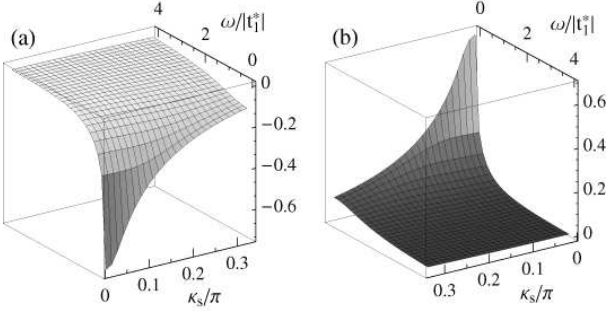}
}
\caption[14]{ 
(a) Nearest neighbor  $\mbox{Re}\bigl[X_1(\omega)\bigr]$ and
(b) next-nearest neighbor
$\mbox{Re}\bigl[X_2(\omega)\bigr]$. 
Note that the directions of $\kappa_s$ and $\omega$ axes are
opposite to each other between (a) and (b).  }
\label{Fig_exch2}
\end{figure} 

Cooper pairs of the $d\gamma$ wave are mainly bound by the enhanced
exchange interaction.  It is divided into the {\em bare} one
and the enhanced part, as is shown in Eq.~(\ref{EqIs*2B}). As is
shown in Eq.~(\ref{EqIs1}), the {\em bare} one is further divided
into three terms.  The first term is the superexchange interaction.
It is antiferromagnetic, and it is as large as 
\begin{equation}\label{J-1}
|J/t_1^*| \simeq (2\mbox{--}3). 
\end{equation}
The second term is the exchange interaction arising from the
virtual exchange of pair excitations of quasi-particles:
\begin{equation}
\frac1{4}J_Q(\omega+i0,{\bf q}) =
\frac{2\tilde{W}_s^2}{\tilde{\chi}_s^2(0)}
\sum_{i=1}^{+\infty} P_i(\omega+i0)\eta_{is}({\bf q}) ,
\end{equation}
with
\begin{equation}
P_i(\omega+i0) =
\frac1{N}\sum_{\bf q} P (\omega+i0,{\bf q})\eta_{is}({\bf q}). 
\end{equation}
Fig.~\ref{Fig_exch1}(a) shows nearest-neighbor 
$\mbox{Re}\bigl[P_1(\omega+i0)\bigr]$, and  Fig.~\ref{Fig_exch1}(b)
shows next-nearest-neighbor 
$\mbox{Re}\bigl[P_2(\omega+i0)\bigr]$. The nearest-neighbor
one is antiferromagnetic for almost half-filling and is
stronger as the electron density is closer to  half-filling.
However, it is less effective than the superexchange interaction:
\begin{equation}\label{J-2}
\frac{2\tilde{W}_s^2}{\tilde{\chi}_s^2(0)}
\frac{|P_1(+i0)|}{|t_1^*|} \alt 0.1. 
\end{equation} 
It is interesting that $\mbox{Re}\bigl[P_1(\omega+i0)\bigr]$ is
ferromagnetic for $n<0.7$. The ferromagnetic exchange interaction
must play a role in the reduction of $T_c$ in the over-doped region.
 The third term is  the so called mode-mode coupling term
$-\Lambda(\omega+i0,{\bf q})$. This term works as an
repulsive interaction for $d\gamma$-wave Cooper pairs. 
The enhanced part of Eq.~(\ref{EqIs*2B}) is approximately given by
\begin{equation}\label{EqEnI}
\mbox{$\left[\frac1{4}I_s(0,{\bf Q})\right]^2\chi_s 
(\omega +i0,{\bf q})$},
\end{equation}
with
\begin{eqnarray}
&&\hspace*{-1cm}
\chi_s (\omega + i0,{\bf q}) =
\chi_s(0,{\bf Q}) \kappa_s^2
\sum_{i=0}^{+\infty} X_i(\omega+i0) \eta_{is}({\bf q}), 
\\ &&\hspace*{-0.5cm}
X_i(\omega+i0) = \frac1{N} \sum_{\bf q}
\frac{\eta_{is}({\bf q})}
{\kappa_s^2 + (qa)^2 - i \omega/\Gamma_s}.
\end{eqnarray}
Fig.~\ref{Fig_exch2} shows nearest-neighbor 
$\mbox{Re}\bigl[X_1(\omega+i0)\bigr]$  and  next-nearest-neighbor 
$\mbox{Re}\bigl[X_2(\omega+i0)\bigr]$. The nearest-neighbor
one is also antiferromagnetic. As the electron density is closer to
half-filling, $\kappa_s$'s become smaller so that the enhanced part 
is stronger. It is also less effective than the superexchange
interaction:
\begin{equation}\label{J-3}
\left[\frac1{4}I_s(0,{\bf Q} )\right]^2\!
\frac{\chi_s(0,{\bf Q}) \kappa_s^2}{|t_1^*|}
|\mbox{Re}[X_1(\omega \!+\! i0)]| \alt 0.5 ,
\end{equation}
for $|\omega| \simeq k_BT_K$. We can argue from Eqs.~(\ref{J-1}),
(\ref{J-2}) and (\ref{J-3}) that the superexchange interaction is
the main part of the pairing interaction in high-$T_c$ cuprate
oxides, and that the total pairing interaction is larger as the
electron density is closer to half-filling. 

In the weak-coupling Hubbard model, the spin-fluctuation
mediated pairing interaction is given by
\begin{equation}\label{EqUChi} 
U^2 \chi_s(\omega+i0,{\bf q}) .
\end{equation}  
Not a few people claim that Cooper pairs must be bound by
Eq.~(\ref{EqUChi}) in high-$T_c$ cuprate oxides.  However, this 
pairing mechanism cannot apply to the cuprate oxides, which certainly
lie in the strong-coupling regime.  Eq.~(\ref{EqUChi}) relevant in the
weak-coupling regime is certainly smoothly connected with the pairing
mechanism by  Eq.~(\ref{EqIs*1}) or (\ref{EqIs*2}) relevant in the
strong-coupling regime.  However, Eq.~(\ref{EqUChi}) is physically
different from Eqs.~(\ref{EqIs*1}) and (\ref{EqEnI}), even if it is
similar to them in appearance.  Because Eq.~(\ref{EqUChi}) arises
from the virtual exchange of low-energy pair excitations of
quasi-particles, it can also be called an exchange interaction. If we
call the superexchange interaction a spin-fluctuation mediated
interaction, on the other hand, it sounds impertinent because the
energy scale of spin fluctuations responsible for the superexchange
interactions is as large as the Hubbard repulsion $U$ and is much
larger than the effective Fermi energy of quasi-particles.

\section{Conclusion}
\label{conclusion}

A theory of Kondo lattices is developed for the $t$-$J$ model.  On
the basis of the microscopic theory, a phenomenological theory of
high-$T_c$ superconductivity in cuprate oxides is developed; 
parameters  are phenomenologically determined instead of completing
the self-consistent procedures involved in the microscopic theory. 

In the strong-coupling regime, electrons are mainly renormalized by
local quantum spin fluctuations so that the density of states is of a
three-peak structure, Gutzwiller's quasi-particle band at the
chemical potential between the lower and upper Hubbard bands.  Two
{\em bare} exchange interactions between quasi-particles are
relevant: One is the superexchange interaction, which arises from the
virtual exchange of spin-channel pair excitations of electrons
between the lower and upper Hubbard bands, and  the other is the
exchange interaction arising from that of spin-channel pair
excitations of quasi-particles themselves. Gutzwiller's
quasi-particles are further renormalized by antiferromagnetic and
superconducting fluctuations caused by intersite exchange
interactions. The sum of {\em bare} exchange interactions including
the two are enhanced by low-energy antiferromagnetic fluctuations
into the enhanced one. The condensation of
$d\gamma$-wave Cooper pairs bound by the enhanced exchange
interaction is responsible for high-$T_c$ superconductivity in the
optimal-doped region. 

In the under-doped region, inelastic scatterings by $d\gamma$-wave
superconducting low-energy fluctuations cause large life-time widths
of quasi-particles around X points; they are almost linear or
sublinear in $T$. Anisotropic pseudo-gaps open because spectral
intensities of quasi-particles around X points are swept away  by
strong inelastic scatterings. The development of pseudo-gaps with
decreasing $T$ cannot be monotonic; their magnitude must decrease at
temperatures close to  $T_c$. Superconductivity eventually occurs
when the pair breaking by inelastic scatterings becomes small enough 
at low enough $T$'s. Superconducting gaps $\epsilon_G(0)$ at $T=0$~K
are never much reduced, because low-energy antiferromagnetic and
superconducting fluctuations are substantially suppressed by the
opening of $\epsilon_G(0)$'s themselves and, in particular, because
their thermal fluctuations vanish at $T=0$~K.

\begin{acknowledgments}
The author is thankful to M. Ido, M. Oda and N. Momono for useful
discussion and suggestion. This work was supported by a Grant-in-Aid
for Scientific Research (C) No.~13640342 from the Ministry of
Education, Cultures, Sports, Science and Technology of Japan.
\end{acknowledgments}

\begin{widetext}
\appendix*
\section{}
 
The summations over $\varepsilon_n$ in Eqs.~(\ref{EqP}) and
(\ref{EqIrredPiSC}) can be  analytically carried out. When the
analytical continuation such as $i\omega_l \rightarrow \omega +i0$ is
made, they are given by 
\begin{equation}\label{EqPA}
P(\omega+i0,{\bf q}) = 
\frac1{i \pi N} \sum_{{\bf k}}
R[\omega, \xi({\bf k}+{\bf q}), \xi({\bf k})], 
\end{equation}
\begin{equation}\label{EqIrredPiSCA}
\pi_{d\gamma}^*(\omega+i0,{\bf q}) =
\frac1{i 2\pi N} \sum_{\bf k}
\eta_{d\gamma}^2({\bf k})
R\left[
\omega, \xi\left({\bf k}+\mbox{$\frac1{2}$}{\bf q}\right),
- \xi\left(-{\bf k}+\mbox{$\frac1{2}$}{\bf q}\right)
\right] ,
\end{equation}
with
\begin{eqnarray}
R(\omega,\xi_1,\xi_2) &=&
\frac1{\omega-\xi_1+\xi_2} \left[ -\psi\left(
\frac1{2}+\frac{-i\omega+i\xi_1 + \gamma}{2\pi k_BT}\right)
+\psi\left( \frac1{2}+\frac{i\xi_2 + \gamma}{2\pi k_BT }\right)
\right.
\nonumber\\ && 
\left. \hspace*{1cm} -\psi\left( \frac1{2} + \frac{-i\omega-i\xi_2+\gamma}{2\pi k_BT}\right) +\psi\left( \frac1{2} + \frac{-i\xi_1+\gamma}{2\pi k_BT}\right)
\right]
\nonumber\\ &&  +\frac1{\omega-\xi_1+\xi_2 +i 2
\gamma} 
\left[
\psi \left(\frac1{2}+ \frac{-i\omega+i\xi_1 + \gamma}
{2\pi k_BT} \right) - \psi \left(\frac1{2}+ 
\frac{-i\xi_2 + \gamma}{2\pi k_BT} \right)
\right.
\nonumber\\ && 
\left. \hspace*{1cm} + \psi \left(\frac1{2}+
\frac{-i\omega-i\xi_2 + \gamma}{2\pi k_BT} \right) -\psi
\left(\frac1{2}+ \frac{i\xi_1 + \gamma}{2\pi k_BT} \right)
\right] ,
\end{eqnarray}

\end{widetext}


\begin{thebibliography}{}
\bibitem{comHubGut}
The configuration of up- and down-spin electrons fluctuates. A
characteristic energy scale of the fluctuations is $k_BT_K$. When
high-energy single-particle excitations such as 
$|\varepsilon|\gg k_BT_K$ are considered, we can regard the
configuration  as being almost static. The problem is reduced to one
for static disordered systems, which can be treated in the coherent
potential approximation (CPA). Hubbard's is equivalent to CPA and is
relevant for $|\varepsilon|\gg k_BT_K$.  On the other hand,
Gutzwiller's is a variational theory and is relevant for 
$|\varepsilon|\alt k_BT_K$.
\bibitem{SSAcom}
Single-site terms should be renormalized by intersite effects in an
eventual theory.
Then, various SSA's are possible depending on the renormalization.
\bibitem{vertex}
Multi-site vertex corrections consistent with multi-site self-energy
corrections should be considered to satisfy the Ward
relation.\cite{ward}
\end{thebibliography}
\end{document}